\newcommand{\CLASSINPUTtoptextmargin}{0.9in}
\newcommand{\CLASSINPUTbottomtextmargin}{1in}
\documentclass[lettersize,journal]{IEEEtran}
\IEEEoverridecommandlockouts

\usepackage{amsfonts}   
\usepackage{amssymb}  
\usepackage{amsmath}  
\usepackage{stmaryrd}
\SetSymbolFont{stmry}{bold}{U}{stmry}{m}{n}

\usepackage{amsthm}
\usepackage{mathtools}
\usepackage[dvipsnames]{xcolor}
\usepackage[locale=US]{siunitx}
\usepackage{graphicx}
\usepackage{anyfontsize}
\usepackage[numbers,sort&compress]{natbib}
\def\BibTeX{{\rm B\kern-.05em{\sc i\kern-.025em b}\kern-.08em
T\kern-.1667em\lower.7ex\hbox{E}\kern-.125emX}}
\usepackage{enumitem}
\usepackage{soul}
\usepackage{stmaryrd}
\usepackage{bm}
\usepackage{multirow}
\usepackage[aboveskip=2pt]{subcaption}
\usepackage{lipsum}
\usepackage{algorithm}
\usepackage[noend]{algpseudocode}
\usepackage{tabularx}
\usepackage{diagbox}
\usepackage{tabularray}
\usepackage{onimage}            
\usepackage{tikz}   
\usepackage{setspace}
\UseTblrLibrary{booktabs}

\makeatletter
\newcommand{\raisemath}[1]{\mathpalette{\raisem@th{#1}}}
\newcommand{\raisem@th}[3]{\raisebox{#1}{$#2#3$}}
\makeatother

\PassOptionsToPackage{hyphens}{url}\usepackage{hyperref}

\input{mysymbol.sty}
\newtheorem{problem}{Problem}
\title{
GLANCE: Graph-based {Learnable} Digital Twin for Communication Networks
    \thanks{%
        Research was sponsored by the Army Research Office under Cooperative Agreement Number W911NF-19-2-0269 and partially supported by USA NSF under Award CCF-2008555.
        The views and conclusions contained in this document are those of the authors and should not be interpreted as representing the official policies, either expressed or implied, of the Army Research Office or the U.S. Government. 
        The U.S. Government is authorized to reproduce and distribute reprints for Government purposes notwithstanding any copyright notation herein.
        E-mails: \{boning.li, tde1, ak109, segarra\}@rice.edu,         gunjan.verma.civ@army.mil. 
    }
}

\author{\IEEEauthorblockN{
Boning Li\IEEEauthorrefmark{1},
Gunjan Verma\IEEEauthorrefmark{2}, 
Timofey Efimov\IEEEauthorrefmark{1}, 
Abhishek Kumar\IEEEauthorrefmark{1},
and
Santiago Segarra\IEEEauthorrefmark{1}}\\
\IEEEauthorblockA{
        \IEEEauthorrefmark{1}Rice University, USA
        \hspace{1em}
        \IEEEauthorrefmark{2}DEVCOM Army Research Laboratory, USA
    }
}

\begin{document}
\setlength{\abovedisplayskip}{3pt}
\setlength{\belowdisplayskip}{3pt}

\maketitle

\begin{abstract}

As digital twins (DTs) to physical communication systems, network simulators can aid the design and deployment of communication networks.
However, time-consuming simulations must be run for every new set of network configurations. 
Learnable digital twins (LDTs), in contrast, can be trained offline to emulate simulation outcomes and serve as a more efficient alternative {to simulation-based DTs} at runtime.
In this work, we propose GLANCE, a communication LDT that learns from the simulator \mbox{ns-3}.
It can evaluate network key performance indicators (KPIs) and assist in network management with exceptional efficiency.
Leveraging graph learning, we exploit network data characteristics and devise a specialized architecture to embed sequential and topological features of traffic flows within the network. 
In addition, multi-task learning (MTL) and transfer learning (TL) are leveraged to enhance GLANCE's generalizability to unseen inputs and efficacy across different tasks.
Beyond end-to-end KPI prediction, GLANCE can be deployed within an optimization framework for network management.
It serves as an efficient or differentiable evaluator in optimizing network configurations such as traffic loads and flow destinations.
Through numerical experiments and benchmarking, we verify the effectiveness of the proposed LDT architecture, demonstrate its robust generalization to various inputs, and showcase its efficacy in network management applications.
\end{abstract}


\section{Introduction}\label{s:intro}

Researchers and engineers use digital twins (DTs) to predict the performance of physical systems and to improve their design before actual implementation~\cite{masaracchia2022digital}.
By providing an economical and safe means to virtually model the physical world, DTs have found success across various industries such as manufacturing, healthcare, urban planning, and communications~\cite{kritzinger2018digital,elayan2021digital,liao2021digital,zhao2020intelligent}. 
Particularly in communication systems, network DTs have gained significant traction due to the ever-growing scale and complexity of modern networks and wireless technologies~\cite{li2023adaptive, zhang2023digital, khan2022digital}.
{For example, discrete-event simulators} like \mbox{ns-3}~\cite{henderson2008network,carneiro2010ns} or OMNeT++~\cite{varga2010overview} are software tools programmed to create virtual network components (e.g., routers, hosts, and channels) and simulate data transmissions among them based on prespecified configurations, protocols, and user-defined scenarios.
{A number of factors contribute to slow and computationally intensive simulations, including the comprehensive profiling of various activities~\cite{arvind2016comparative},}
intricate dependencies among network components, and the need to solve sophisticated equations with iterative algorithms~\cite{campanile2020computer,fujimoto2022network}.
{Moreover, their complexity scales with the number of events (or data packets, specific to this context), which grows rapidly with the number of nodes, flows, or link speed.}
Hence, simulation-based DTs face a fatal efficiency disadvantage when dealing with complex systems. 
{In contrast, learnable digital twins (LDTs) facilitate higher efficiency by learning} an end-to-end mapping from network configurations to {their resulting} performance~\cite{ye2021deep,zheng2023end}.
{In the absence of a physical system to provide abundant training data, simulators can bridge the gap~\cite{hu2023simulation}.}
While {simulating} training data and completing the training may consume considerable time and resources, these processes need to be performed offline only once.
Upon deployment, {the LDT} simply executes a feedforward pass to {infer} KPIs or other outcomes trained on, which is {easily three or more orders of magnitude faster than a simulator}~\cite{li2023learnable}. 

Various designs of learnable architectures are plausible for network DTs, so long as they align with the specific problem, data, and objective~\cite{rusek2020routenet,ferriol2022routenet,ferriol2023routenet,kumar2018novel,lin2021stochastic,poularakis2021generalizable,bellavista2021application,sun2019gradientflow}.
As a series of seminal works, RouteNet and its variants~\cite{rusek2020routenet,ferriol2022routenet,ferriol2023routenet} learned from sequential links within individual transmission paths and shared links across multiple paths.
These specialized recurrent and relational architectures proved to be more effective in handling multiple flows' coexistence in wired networks compared to the analytical modeling by queuing theory. 
Yet, {their performance falls short in wireless scenarios}, where unaccounted node interference can significantly influence wireless performance~\cite{jain2003impact}. 
In addition, node mobility causes time-varying channels and requires the handling of different instantaneous network topologies~\cite{an2023ml}.
Graph neural networks (GNNs) are valued for their ability to model node interference and interactions as graph structures, making them versatile tools in various wireless contexts, including beamforming ~\cite{chowdhury2023deep,li2024gnn}, link scheduling~\cite{zhao2022link}, and resource allocation~\cite{chowdhury2021unfolding,li2022graph,shen2022graph}.
However, {amid the use of GNNs, LDT architectures are rarely specialized to explicitly represent wireless dynamics, as evidenced in recent surveys~\cite{isah2024graph,almasan2022digital}.}
Graph-learning LDTs hold great untapped potential in large and dynamic wireless networks {with behaviors difficult} to model or predict, and hence optimize a priori.

Network LDTs' utility extends beyond network evaluation alone. 
They can serve as integral components within larger applications where efficient evaluation plays a crucial role.
For example, in a power grid DT,~\cite{danilczyk2021smart} introduced a convolutional neural network (CNN) module that can scan the bus voltage data for anomalies, enabling real-time fault detection when running alongside the physical grid.
Additionally,~\cite{thomas2024digital} proposed an online-learning-based DT to estimate the capacity region of a wireless ad hoc network, in order to support a working wireless network testbed.
Furthermore, numerous works have employed LDTs in reinforcement learning (RL)~\cite{tang2023digital,deng2021digital,10089851,zhang2023digital_JSAC,zhang2021adaptive,zhao2023graph}.
In actor-critic RL, LDTs can act as the critic network to assess the environment and evaluate selected actions. 

Taking all into account, we draw inspiration from both RouteNet and GNNs to craft a novel LDT architecture tailored to our specific scenario, namely wired or wireless networks with the coexistence and potential interference of multiple traffic flows.
On certain occasions, one may desire an all-encompassing LDT capable of predicting multiple KPIs.
A straightforward option is to train and ensemble of separate LDTs for different KPIs, known as single-task learning (STL).
{However, STL has main shortcomings in its limited ability to transfer knowledge between tasks, leading to suboptimal performance and complicated model maintenance~\cite{zhang2021survey,zhuang2020comprehensive}.}
To tackle these challenges, we explore the paradigms of multi-task learning (MTL) and transfer learning (TL), which are also widely used techniques for LDTs~\cite{xu2019digital,wang2023digital,lu2021adaptive,ma2024driver}.  
Additionally, we set up network management applications, such as traffic load optimization~\cite{hassanein2001routing} {and flow destination selection}~\cite{kumar2013survey}, where we can further validate the effectiveness of the trained LDT.

\vspace*{2mm}
\noindent\textbf{Paper outline. }
We propose a novel Graph-enhanced Learning Architecture featuring Network embeddings for Communication network Evaluation, acronymed GLANCE.
In Section~\ref{s:pre}, we formulate the learnable twinning {of a physical communication system as a supervised problem of learning the mapping between input and output of \mbox{ns-3}.}
In Section~\ref{s:alg:arch}, we elaborate on the proposed architecture and algorithm.
Our training strategies are laid out in Section~\ref{ss:learning_strategies} and applications are discussed in Section~\ref{ss:manage}.
Moving on to Section~\ref{s:data}, we introduce data generation, test benchmarks, training configurations, and other elements to help understand the datasets used in this study.
Numerical experiments and results are presented in Section~\ref{s:exp}.
Finally, we conclude this paper by summarizing the performance and potential of GLANCE in Section~\ref{s:conclusion}.
For important notation, please refer to Table~\ref{tab:notation} in Appendix~\ref{ap:notations}.

\section{Problem Formulation}\label{s:pre}

\begin{figure}[t]
\centering
    \includegraphics[width=\linewidth,trim= 0cm 2.5cm 0 .3cm]{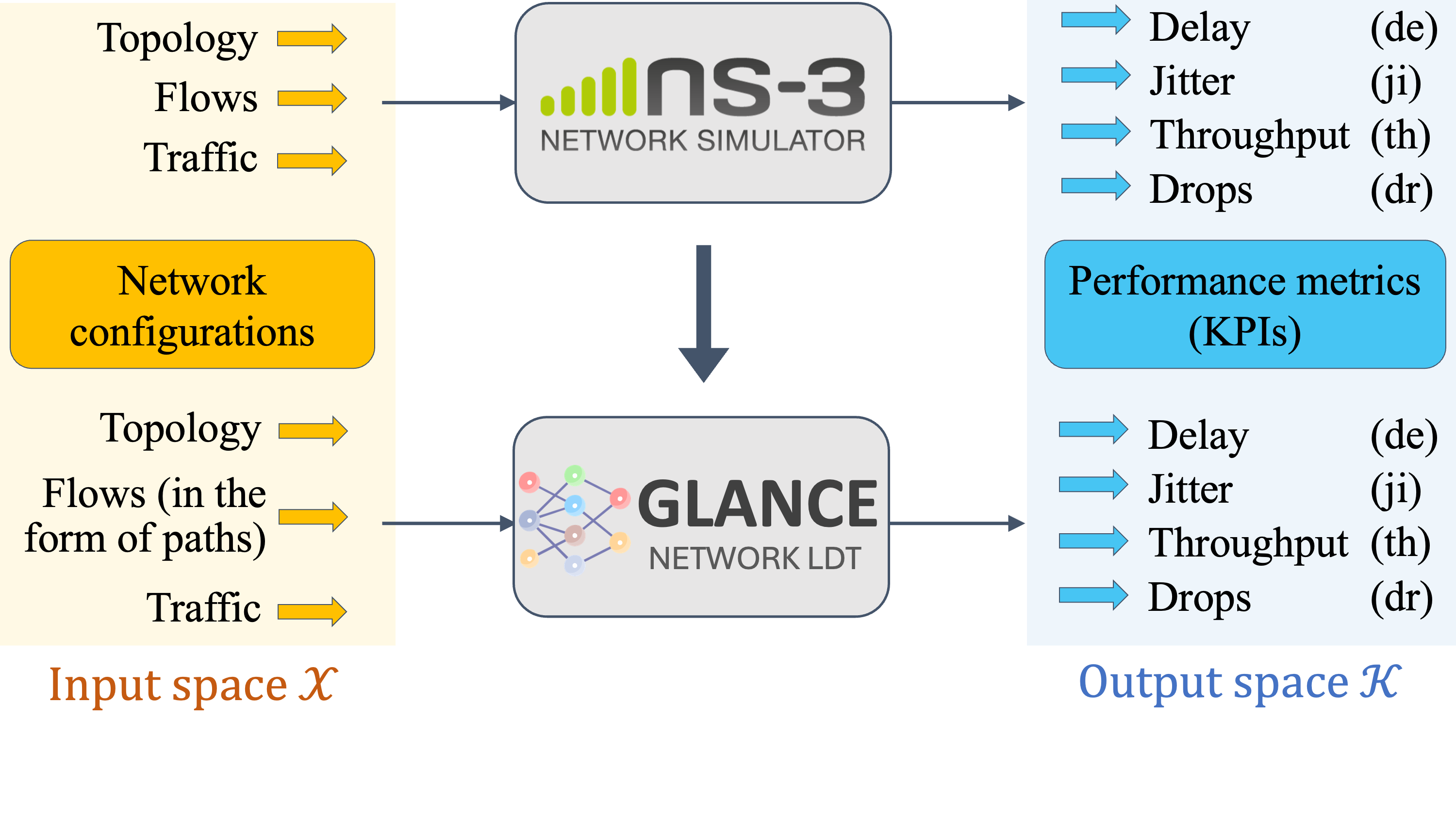}
    \caption{
    {A network simulator's role as a DT in predicting KPIs for a given network can be replaced with a more efficient LDT.}
    }
    \label{f:overview}
\end{figure}

Supervised learning provides a reasonable framework for acquiring {a learning-based (LB) counterpart} of a simulator through data-driven approaches. 
We now introduce the problem formulation for training an LDT, outlining its inputs and outputs as follows.
\begin{problem}
    Consider a communication network system $S{:\,}\ccalX{\,\rightarrow\,}\ccalK$ with configurations in $\ccalX$ and performance metrics in $\ccalK$.
    A parameterized architecture $\Psi{:\,}\ccalX{\,\rightarrow\,}\ccalK$ is devised as a DT to the system, sharing the same input and output spaces as $S$. 
    Find its parameters $\bbW^*$ to minimize a loss function $f$
    \begin{equation}\label{p:twin}
        \bbW^* = \argmin_\bbW \mathop\mbE_{\bbx \sim \ccalD(\ccalX)}\Big[f\left(\Psi(\bbx;\bbW), S(\bbx)\right)\Big],\tag{P1}
    \end{equation}
    over a distribution of inputs $\ccalD(\ccalX)$.
\end{problem}
{To clarify, the system $S$ is ideally a physical network $S_\mathrm{phy}$ but can be practically substituted by a DT simulator $S_\mathrm{sim}$ like \mbox{ns-3} that approximates its behaviors.}
A schematic view is presented in Fig.~\ref{f:overview}.
For our specific application of interest, the input space $\ccalX$ encompasses three sub-spaces: the network topology ($\ccalX_g$), the predefined flows ($\ccalX_f$), and the traffic input ($\ccalX_t$).
{The output space $\ccalK$ represents the types of KPIs of interest.}

\subsection{Input: Network configurations}

The topology can be modeled as a directed graph $\ccalG{\,=\,}(\ccalN, \ccalL, \bbA)$.
Here, $\ccalN$ denotes the set of $N$ nodes.
The link set $\ccalL{\,\subseteq\,}\{(i,j){\,|\,}i, j{\,\in\,}\ccalN, \text{and } i{\,\neq\,}j\}$ is a subset of all possible ordered pairs of nodes such that $(i, j){\,\in\,}\ccalL$ if and only if $i$ can transmit data directly to $j$ without relays.
The adjacency matrix $\bbA{\,\in\,}\mbR^{N\times N}$ satisfies $A_{ii}{\,=\,}0$, $A_{ij}{\,>\,}0$ if $(i, j){\,\in\,}\ccalL$ and $A_{ij}{\,=\,}0$ otherwise. 
Additionally, the degree matrix $\bbD{\,\in\,}\mbR^{N\times N}$ is a diagonal matrix with $D_{ii}{\,=\,}\sum_{j}(A_{ij}{+}A_{ji})/2$ denoting the degree of node $i$. 
For simplification, we implement symmetric {path} losses for all simulations, reducing the need to consider link directions in the graph-learning context. 
In wired networks, because of their relatively reliable transmission characteristics, links are often considered unweighted.
However, in wireless networks, where path loss depends on node distance and interference can heavily influence performance, assigning non-trivial weights to links becomes important. 
These weights can represent signal strength or other relevant metrics, helping characterize the quality of wireless connections and aiding in network emulation and optimization.

Data are generated by applications in source nodes and transmitted to destination nodes, creating traffic flows in the network. 
Traffic can be of various patterns and characteristics, often modeled as random variables (RVs). 
In our scenario, we utilize an on/off pattern that keeps switching between `on' (when data are generated) and `off' (when the generation is paused) phases. 
{For each flow,} the `on' and `off' durations are sampled from random distributions upon switches, i.e., {$t_{\on,i,j}{\,\sim\,}\ccalT(\tau_{\on,i})$ and $t_{\off,i,j}{\,\sim\,}\ccalT(\tau_{\off,i})$ for the $i^\mathrm{th}$ flow's $j^{\mathrm{th}}$ on and off switches, respectively.}
{To further clarify,} while DTs' interfaces are provided with the parameters $\bbtau_{\{\on,\off\}}$ defining $\ccalT$, they do \emph{not} possess knowledge of the specific sampled $\bbt_{\{\on,\off\}}$ values during the simulation. 
In other words, DTs operate based on statistical information rather than exact values of the simulated traffic.
This is an important source of stochasticity within the system.

A flow is defined as a pair of source and destination nodes without considering the precise route taken.
Let us denote the set of $F$ flows as $\ccalF{\,=\,}\{(s_i,d_i){\,|\,}\forall i{\,<\,}F\}$, or alternatively as two vectors {$\bbf_{\{\src,\dst\}}{\,\in\,}\ccalN^F$, where $\bbf_\src[i]{\,=\,}s_i$ and $\bbf_\dst[i]{\,=\,}d_i$, i.e.,} each containing source and destination indices, respectively. 
A path, or an ordered sequence of links, directs how data packets travel within the network to reach their intended destination.
It is defined as a sequence of consecutive links going from the source node to the destination node of a flow. 
Collectively, the set of paths of all flows in a network is denoted as $P(\ccalF)$. 
To find to $P(\ccalF)$, a routing protocol must be followed.
One such protocol {commonly built-in for simulators} is optimized link state routing (OLSR)~\cite{jacquet2001optimized}, a proactive and distributed regime using Dijkstra's shortest path algorithm for path computation. 
However, GLANCE is not equipped with OLSR functions. 
Although one can {apply the Dijkstra's algorithm as a preprocessing step of LB twinning, these externally found paths} may not precisely match those used during simulation, especially if multiple shortest paths are available for a flow.
To ensure accurate path information for GLANCE, we keep a record of the routing tables that \mbox{ns-3} determined and used during each simulation.
Instead of flows, we provide paths to GLANCE, and these paths are identical to those used in the simulation, thereby enhancing the twinning accuracy.
Section~\ref{s:data:sim} will provide more details about the simulation process within \mbox{ns-3}.\looseness=-1

\subsection{Output: Network KPIs}\label{sss:output}
In \mbox{ns-3}, KPIs are estimated for each flow by monitoring statistics related to the simulated activities over a prespecified duration.
These KPIs include average delay, jitter, throughput, and packet drops, as indicated in Fig.~\ref{f:overview}.
Focused on different perspectives such as latencies, stability, and reliability, they can collectively offer comprehensive insights into the network's performance.

In summary, easy, fast, and accurate prediction of network KPIs is crucial for network engineers to proactively detect network congestion, identify faulty components, and optimize network resources, all of which contribute to maintaining optimal network performance and delivering a satisfactory user experience.

\section{Methods}\label{s:alg}

We propose GLANCE as an {efficient LB approximation to ns-3 in the role of a DT for predicting network KPIs.}
Upon completion of training, it can empower network engineers with swift and accurate evaluation capabilities, based on which network management decisions such as traffic loads and destinations can be optimized efficiently.

\subsection{Architecture and algorithm}\label{s:alg:arch}

\begin{figure}[t]
\centering
    \includegraphics[width=.9\linewidth,trim= 0 3.9cm 3.7cm .2cm]{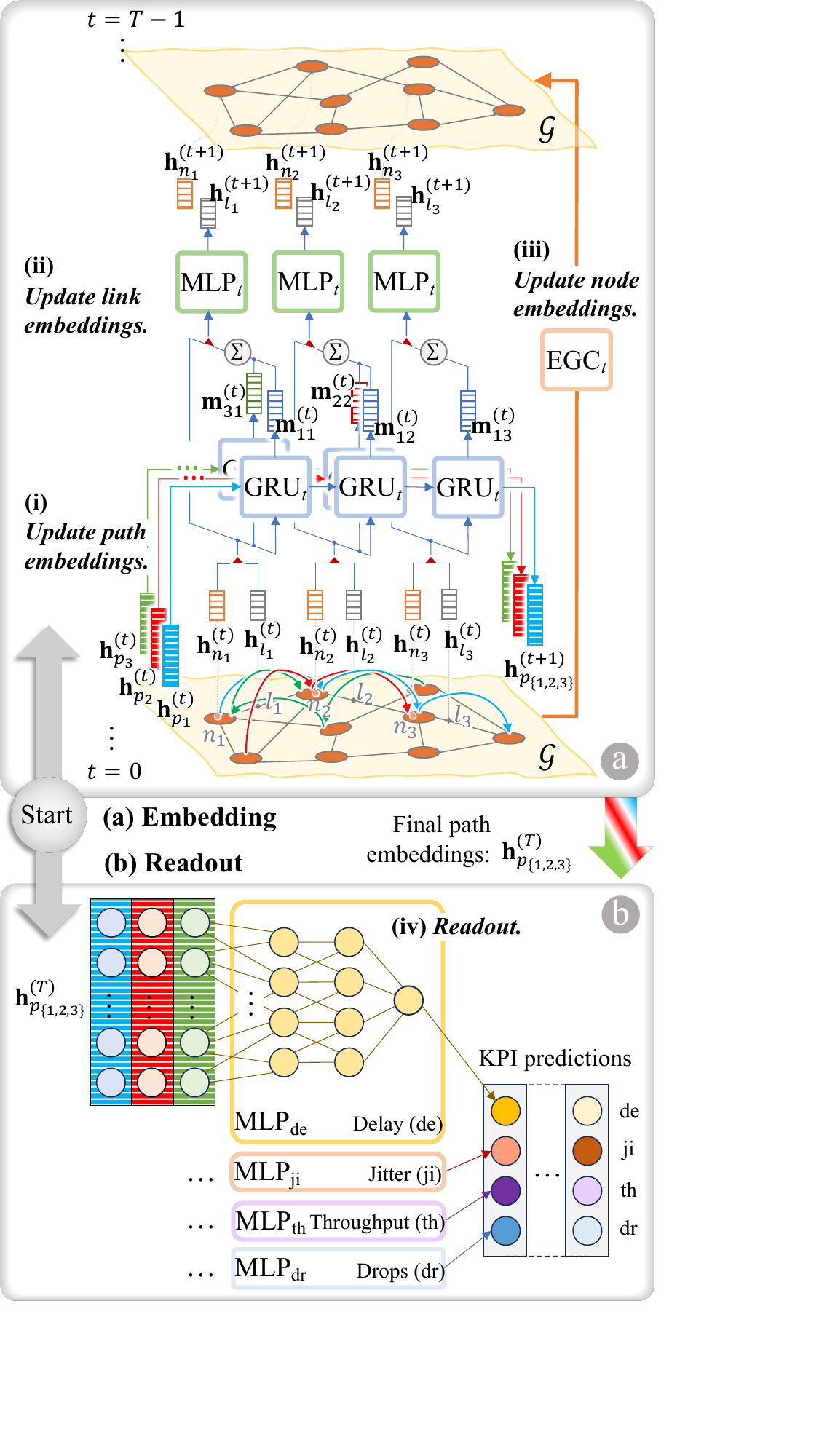}
    \caption{
    GLANCE architecture consists of 
    (a)~an embedding module of $T$ stacked layers and 
    (b)~a readout module of $K$ parallel blocks.
    }
    \label{f:scheme}
\end{figure}

Figure~\ref{f:scheme} illustrates the architecture of GLANCE.
Algorithm~\ref{alg:cap} details each step in its feedforward pass.
At a high-level abstraction, GLANCE consists of two modules: {the first module acquires network embeddings, and the second module translates} these embeddings into KPI predictions. 
Down the hierarchy, the embedding module comprises $T$ GLANCE layers.
These layers are highly specialized and composed of \emph{path}, \emph{link}, and \emph{node} sub-networks.

\begin{algorithm}[t]
\caption{
GLANCE algorithm.
}\label{alg:cap}
\begin{algorithmic}[1]
\Require Graph $\ccalG=(\ccalN, \ccalL, \bbA)$, path list $\bsP=P(\ccalF)$ 
\Ensure Traffic $\bbtau_\onoff$, link capacities $\bbc$, node degrees $\bbD$
\State Path embeddings\label{alg:cap:l1}
$$\bbh_{p_i}^{(0)}\gets  [\tau_{\on,i},\tau_{\off,i}]{\,\mathbin\Vert\,}\bbzero,\text{ with }i\text{ indexing }\forall\, \bbp\in \bsP$$
\State Link embeddings
$$\bbh_{l_r}^{(0)}\gets [c_r]{\,\mathbin\Vert\,}\bbzero,\text{ with }r\text{ indexing }\forall\, \bbl\in\ccalL$$
\State Node embeddings
\label{alg:cap:l2}
$$\bbh_{n_u}^{(0)}\gets [D_{uu}]{\,\mathbin\Vert\,}\bbzero,\text{ with }u\text{ indexing }\,\forall\, n\in\ccalN$$
\For{layer $t= 0, 1, \cdots, T-1$}
\State \textbf{(i)~\textit{Update path embeddings.}}
\For{every path $\bbp$ (indexed by $i$) in $\bsP$}
    \For{every link $\bbl$ (indexed by $s$) in $\bbp$}
        \State With node index $v=\bbl[0]$,
        \State \begin{equation}\label{alg:eq:path}
        \bbh_{p_i}^{(t)}\gets \GRU_t\left(\bbh_{p_i}^{(t)},\, \bbh_{l_s}^{(t)}{\,\mathbin\Big\Vert\,}\bbh_{n_v}^{(t)}\right)\,\,\,\,
        \end{equation}
        \State $\bbm_{i s}^{(t)}\gets \bbh_{p_i}^{(t)}$ at the current GRU step.
    \EndFor
    \State $\bbh_{p_i}^{(t+1)}\gets \bbh_{p_i}^{(t)}$
\EndFor
\State \textbf{(ii)~\textit{Update link embeddings.}}
\For{every link $\bbl$ (indexed by $r$) in $\ccalL$}
    \State With node index $w=\bbl[0]$, and 
    \State path index set $\ccalJ$ indexing $\forall\,\bbp\ni\bbl$,
    \State \begin{equation}\label{alg:eq:link}
    \bbh_{l_r}^{(t+1)}\gets \MLP_t\left(\bbh_{l_r}^{(t)}{\,\mathbin\Big\Vert\,}\bbh_{n_w}^{(t)}{\,\mathbin\Big\Vert\,}\sum\limits_{j}^{\ccalJ}\bbm_{jr}^{(t)}\right)
    \end{equation}
\EndFor
\State \textbf{(iii)~\textit{Update node embeddings.}}
\For{every node $n$ (indexed by $u$) in $\ccalG$}
    \State With link index set $\ccalQ$ indexing $\forall\,\bbl\in L_\mathrm{out}(u)$
    \State  \begin{equation}\label{alg:eq:node}
    \bbh_{n_u}^{(t+1)}\gets \EGC_t\left(\bbh_{n_u}^{(t)}{\,\mathbin\Big\Vert\,}{\sum\limits_q^\ccalQ}{\bbh_{l_q}^{(t)}}; \,\ccalG\right)\quad
    \end{equation}
\EndFor
\EndFor
\State \textbf{(iv)~\textit{Readout.}}
\State $\KPI_k = \MLP_k\left(\bbh_p^{(T)}\right), \,\forall k\in\ccalK$.
\end{algorithmic}
\end{algorithm}

In Algorithm~\ref{alg:cap}, lines~\ref{alg:cap:l1} to~\ref{alg:cap:l2} initialize path, link, and node embeddings as vectors of predefined dimensions.
The operator $\cdot{\,\mathbin\Vert\,}\cdot$ concatenates two vectors into one in the expressed order. 
Initial path, link, and node embeddings encode traffic, link capacity, and node degree information, respectively, in their leading elements. 
The remaining bits are filled with zeros, which notably allow for easy extensibility.
These unused dimensions can potentially incorporate additional network configurations such as the routing protocol, transmit power values, and others.
While these factors are not the primary focus of this study, they may be of interest for future extensions. 

Further elaborating, the path subnet (path-net) consists of gated recurrent unit (GRU) cells~\cite{cho-etal-2014-learning}, a widely used variant of recurrent neural networks (RNNs) for sequential data.
They are suitable because paths are essentially ordered lists of links.
For each path, GRUs integrate the embeddings of the links in the path and their associated nodes based on the order in which packets are transmitted [Algorithm~\ref{alg:cap},~\eqref{alg:eq:path}].
At each recurrent step, intermediate GRU states encapsulate information about the preceding links and nodes in the path.
They are subsequently leveraged to update the corresponding link embeddings.

The link subnet (link-net) contains stacked dense layers, essentially following the structure of multi-layer perceptrons (MLPs).
Since a link may be shared by multiple paths, its influence is not limited to just one path; it may extend to all paths that include this particular link.
Based on this intuition, we need an aggregator, as simple as summation, to combine multiple pieces of information from different paths holistically.
Hence, the input to link-net is a concatenation of link and node embeddings at the previous layer, as well as the sum of path-net's intermediate states across all paths sharing the link.
The output of link-net will be the new embedding of this link, as stated by~\eqref{alg:eq:link} in Algorithm~\ref{alg:cap}. 

The last subnet in a GLANCE layer, namely node-net, features an edge graph convolutional (EGC) layer that processes embeddings of neighboring nodes and links.
It handles interactions, particularly interferences, between nodes during data transmissions.
Its property of being permutation equivariant emphasizes the ability to maintain consistent representations regardless of the indexing of nodes and links, enhancing the robustness and suitability of GLANCE as a network DT.
Through {$T$ stacked GLANCE layers}, the proposed architecture iteratively refines network embeddings, thus capturing the interrelations among network structures.
The final path embeddings are then extracted as the network's representation, serving as the input to the following readout module. 

Regarding the readout module, we maintain individual readout blocks as one-dimensional-output MLPs, one for each type of KPI. 
These non-shared readout blocks {enable GLANCE to accommodate} an arbitrary number of concurrent tasks.
If additional KPIs are introduced at any point, we can easily attach a new readout block for it and seamlessly utilize the new data to learn new predictions.

A limitation of our GLANCE implementation\footnote{GLANCE implementation is in Python/TensorFlow~\cite{tensorflow2015-whitepaper,chollet2015keras,grattarola2021graph} and available at 
\url{https://github.com/bl166/wireless\_digital\_twin\_j}.} 
is the need to specify a maximum path length in accordance with the correct number of recurrent GRU steps for batch training.
{In practice, this may hinder re-training or deployment, should the trained GLANCE be presented with a longer path} (whereas shorter paths are fine with zero-padding).
However, it is not a limitation to Algorithm~\ref{alg:cap}, as GRU and other RNNs, notwithstanding their learning performance, are theoretically capable of taking inputs and outputs of indefinite length.
Removing the path length limitation is planned for future implementation and testing.

{\remark{
The integration of graph learning components as the node-net distinguishes GLANCE from its predecessors and empowers it with the capability to handle wireless dynamics.
}}

\vspace{2mm}
{Concerning this point,~\eqref{alg:eq:node} can be rewritten with the explicit formulation of ECG as follows:}
\begin{equation*}
     \bbh_{n_u}^{(t+1)} = \sigma\left( \hbD^{-\frac{1}{2}}\hbA\hbD^{-\frac{1}{2}} \Big(\bbh_{n_u}^{(t)}{\,\mathbin\Big\Vert}\sum\limits_{l\in L_\mathrm{out}(n)}{\bbh_{l_q}^{(t)}}\Big) \bbW_n^{(t)}\right),
\end{equation*}
where $\hbA$ is the adjacency matrix of $\ccalG$ with added self-loops and $\hbD$ is the corresponding degree matrix for normalization. 
For each node $u$, we aggregate the embeddings of the links emanating from $u$. 
This aggregated link embedding is then concatenated with $u$'s node embedding. 
In addition, $\bbW_n^{(t)}$ is the node-net's weight matrix in the $t^\text{th}$ layer, and $\sigma(\cdot)$ is a non-linear activation function like the ReLU. 
Besides our current choice, there are many other ways to aggregate link embeddings and integrate them with node embeddings in a more sophisticated or learnable manner, including edge-conditioned graph convolutions~\cite{simonovsky2017dynamic}, dynamic edge graph convolutions~\cite{wang2019dynamic}, and so on.
In any case, the purpose of node-net is to incorporate information about neighboring nodes and links, including those from different paths that may interfere with each other, particularly in wireless networks.
{Despite node-net being a simple and commonly used variant of GNN, it plays a crucial role in the specialized architecture of GLANCE.
Holistically, GLANCE's iterative embedding module can capture complex dependencies within communication network systems even under challenging conditions such as wireless interference, high congestion, and topological perturbations.
The effectiveness of this architecture is demonstrated through ablation studies} in Section~\ref{s:exp}. 

\subsection{Multi-task and transfer learning}\label{ss:learning_strategies}

\begin{figure*}[t]
\centering
    \includegraphics[width=\linewidth, trim=0cm 15.3cm 0 0.5cm]{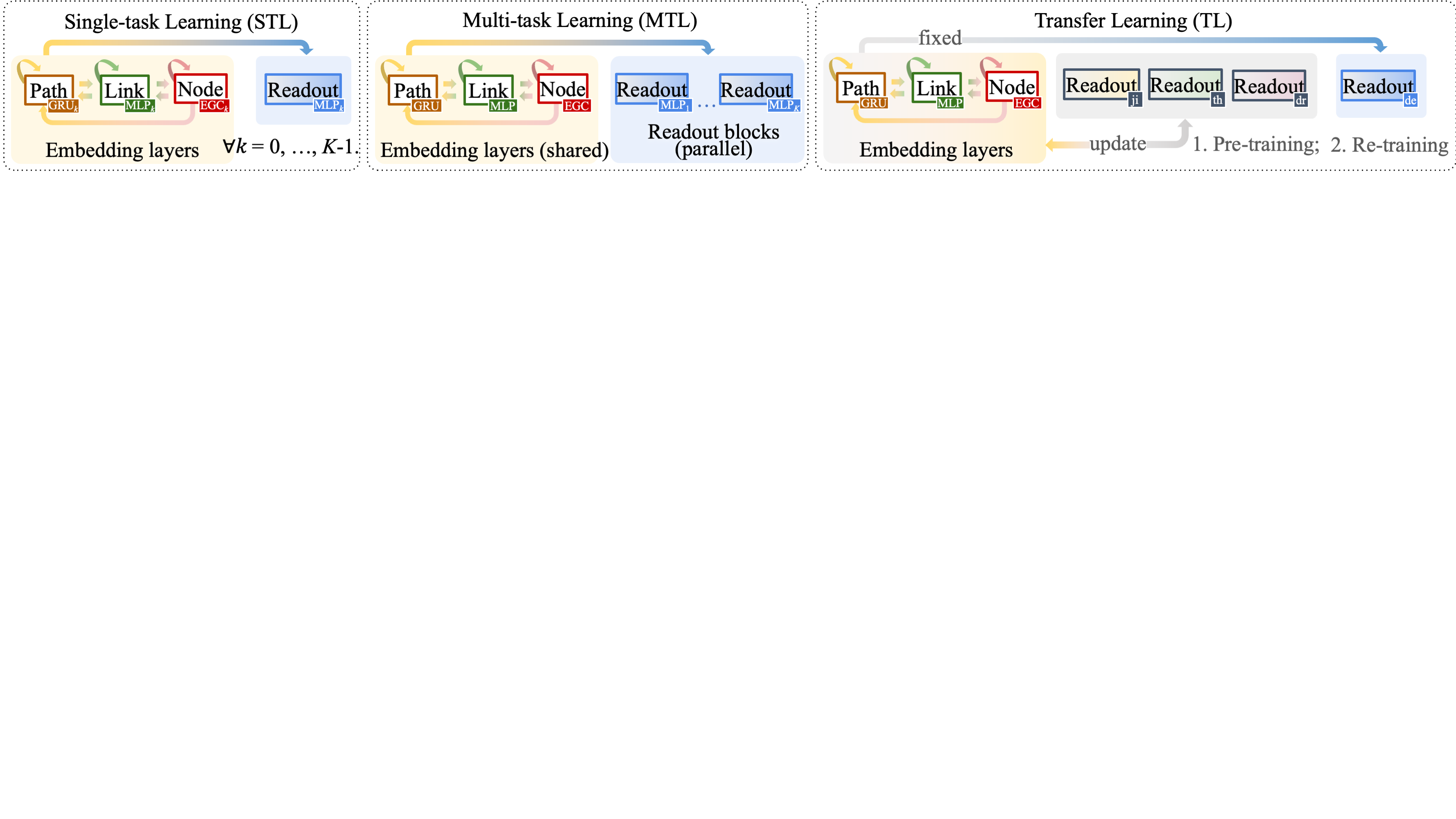}
    \caption{
    Training strategies for GLANCE illustrated.
    }
    \label{f:tr_strat}
\vspace*{-1em}
\end{figure*}

Several supervision strategies can be adopted to train GLANCE. 
Depending on how different tasks (i.e., KPI types) are handled when of concurrent interest, the strategies can be categorized into single-task, multi-task, or transfer learning, as illustrated in Fig.~\ref{f:tr_strat} and briefly discussed below.

\begin{enumerate}[wide,label={\arabic*)},labelindent=0pt]
    \vspace*{1mm}
    \item {\bf Single-task learning (STL)}: 
    Separate models are trained independently for individual tasks without sharing any parameters. 
    STL may be favored for its specificity in a clear and well-defined application focused on a single primary KPI.
 
    \vspace*{2mm}
    \item {\bf Multi-task learning (MTL)}: 
    A single model jointly learns multiple KPIs simultaneously. 
    To enable $K$-task learning, $K$ separate readout blocks are attached to the same embedding block in parallel, each responsible for learning one KPI, as shown in Fig.~\ref{f:scheme}.
    The embedding parameters are shared across all tasks, whereas the readout parameters are segragated for their respective tasks. 

    \vspace*{2mm}
    \item {\bf Transfer learning (TL)}:
    {Specific to our context,} assuming the target task of $\KPI_k$, TL is conducted in two stages. 
    First, an MTL model is {\it pre-trained} using the complementary subset of KPIs, namely $\ccalK\setminus\KPI_k$.
    Second, with embedding parameters fixed and readout blocks detached, a new readout block is attached to the embedding module and {\it re-trained} using the target $\KPI_k$.
    This regime is particularly useful under data or label limitations.
    In scenarios where target labels are scarce, TL offers a viable approach to exploit existing labels, even if they are not specific to the final target.
    By combining MTL and STL ideas, TL initially learns universal network embeddings and then updates the readout parameters to fit the target task. 
    Provided that the pre-trained embeddings have captured comprehensive network information, subsequent re-training can focus on updating the readout parameters and thus see a faster and smoother convergence.  
    
\end{enumerate}

Overall, the choice of strategy depends heavily on the specific problem being addressed.
One must consider task definition, data availability, and performance expectations to implement and validate a suitable training pipeline.

\subsection{Network management}\label{ss:manage}

GLANCE has greater potential beyond its role as a pure network evaluator, as will be elucidated in the context of network management in this section.
Let us formulate the general network management as an optimization problem in the following.

\begin{problem}
    Consider a communication network system $S$ with input entities in $\ccalX$.
    Given a target profile of outcomes $\bbk$, find a network input $\bbx^*$, such that the resulting outcomes of this system are close to the target by some defined metric $f$:\looseness=-1
    \begin{equation}\label{p:man}
        \bbx^* = \argmin_\bbx f(S(\bbx), \bbk).\tag{P2}
    \end{equation}
\end{problem}
A potential solution to~\eqref{p:man} involves iteratively adjusting the network input $\bbx$ to minimize the discrepancy measure between the actual outcomes $\hbk{\,=\,}S(\bbx)$ and the target profile $\bbk$. 
Upon convergence, the optimized $\bbx^*$ represents a configuration that satisfies the desired performance criteria for the system.
In practice, this optimization may employ gradient descent (GD), local search algorithms, or other techniques tailored to the specific problem and system at hand.\looseness=-1

Specifically for our study, the system is an ns-3 simulator. 
The input consists of a known topology $\ccalG$, 
flows $\ccalF$ 
or routing paths $\bsP$, 
and traffic $\bbtau_\onoff{\,\in\,}\mbR^F$, 
with a focus on managing the latter two entities.
By managing the flows, we prefix $\bbf_\src$, the sources of flows, and try to optimize their {destinations $\bbf_\dst$} for achieving desired KPIs.
By managing the traffic, we refer to the optimization of traffic loads, which are characterized by the on/off parameters herein, across fixed flows within a network.
We choose to study these two applications further due to their prevalence in modern network optimization contexts, such as computational offloading~\cite{zhao2024congestion}.
Also, they represent the use of diverse optimization algorithms.
In Algorithm~\ref{alg:cap}, GLANCE directly incorporates traffic into the initial path embeddings, making it differentiable with respect to traffic and suitable for gradient-based algorithms like GD. 
{Regarding GLANCE's use of flows, however, the selection} of in-path links does not involve gradient calculation.
Flow management therefore necessitates non-gradient-based algorithms like hill-climbing. 
Further details on these solutions will be provided in the sequel.

\subsubsection{Traffic load optimization}

Let us assume a $\bbTheta$-parameterized LDT {denoted by $\Psi_{\traf}$ and} validated to generalize well over a continuous traffic space.
Then, we have $\hbk{\,=\,}\Psi_{\traf}(\bbtau;\bbTheta{\,|\,}\ccalG,\bsP){\,\in\,}\mbR^{F\times K}$ {being the predicted KPIs of $K$ types for $F$ flows}, where 
$\bbtau{\,\in\,}\mbR^{2F}$ represents the flattened traffic input, and $\ccalG$ and $\bsP$ are the other two input entities which are fixed herein.
Notice that we can compute the gradient of $\Psi_{\traf}$ with respect to $\bbtau$, denoted as $\nabla_{\tau}\Psi_{\traf}(\bbtau)$, omitting fixed entities for simplicity of notation.
Using a loss function $J(\bbk,\hbk)$, the GD algorithm can be applied to iteratively update $\bbtau$, such as
\begin{equation}\label{e:gd}
    \bbtau_{j+1} = \lfloor \bbtau_j - \alpha \nabla_{\tau}J(\bbk,\Psi_{\traf}(\bbtau_j))\rfloor_0 
\end{equation}
until convergence, with a fixed learning rate $\alpha$. 
The application of a floor function in~\eqref{e:gd} is because of the non-negative nature of traffic, which can potentially be any projection depending on the variable's physical interpretation or other constraints.

\subsubsection{Flow destination selection}

{Let us use $\bbp{\,=\,}\ccalP(s,d)$ to denote the shortest path found for a single flow $(s,d)$, and $\bsP{\,=\,}\ccalP(\bbs,\bbd)$ the vectorized version of a path list for multiple flows with source nodes $\bbs$ and destinations nodes $\bbd$, pairwise.}
Similar to the previous case, we use a trained GLANCE model to predict KPIs for any paths, i.e., $\hbk{\,=\,}\Psi_\pth(\bsP;\bbTheta{\,|\,}\ccalG,\bbtau)$.
We look for a descending direction to minimize the loss $J(\bbk,\hbk)$, but this time on a different and non-differentiable variable $\bsP$.
Therefore, we need to employ a non-gradient approach, taking the hill-climbing algorithm for instance, as detailed in Algorithm~\ref{alg:hillc}. 
Being a greedy local search algorithm, hill-climbing iterations are initialized at a random $\bsP$ (associated with outcome $\hbk$), updated to one of its neighbors $\bsP^\prime$ that results in a better $\hbk^\prime$ such that $J(\bbk,\hbk^\prime){\,<\,}J(\bbk,\hbk)$, and repeated until no better neighbors are found. 

\begin{algorithm}[t]
\caption{Hill-climbing for GLANCE-based flow destination selection.}\label{alg:hillc}
\begin{algorithmic}[1]
\Require Network topology $\ccalG{\,=\,}(\ccalN,\ccalL,\bbA)$, source nodes $\bbs$, routing protocol $\ccalP$, and target profile $\bbk$
\Ensure Random destination nodes $\bbd\in\ccalN^F$, 
\State Corresponding path list $\bsP=\ccalP(\bbs, \bbd)$,
\State Random shuffling function $\pi_f$. \Comment{for $F$ elements}\label{alg:hillc:pif}
\For{every flow index $i$ in $\pi_f(F)$} 
    \State $\hbk=\Psi_\pth(\bsP;\bbTheta{\,|\,}\ccalG,\bbtau)$
    \State $\ccalN^\prime=\ccalN\setminus \{s_i,d_i\}$
    \State Random shuffling function $\pi_n$.  \Comment{for $N{-}2$ elements}\label{alg:hillc:pin}
    \For{every node $n$ in $\pi_n(\ccalN^\prime)$}
        \State $\bsP^\prime=\ccalP(s_j,\, n) \bigcup \bsP\setminus\bsP[i]$
        \State $\hbk^\prime=\Psi_\pth(\bsP^{\prime};\bbTheta{\,|\,}\ccalG,\bbtau)$
        \If{ $J(\bbk,\hbk^\prime) < J(\bbk,\hbk)$ }
            \State Update destination node: $d_i\gets n$ 
            \State Update paths: $\bsP\gets \bsP^\prime$
        \EndIf 
    \EndFor 
\EndFor
\Statex \hspace*{-\algorithmicindent}{\bf Output: } Destination nodes $\bbd$ 
\end{algorithmic}
\end{algorithm}

Compared to~\eqref{e:gd}, Algorithm~\ref{alg:hillc} has broader applicability since it can operate without gradients.
However, its time complexity is higher than the gradient-based algorithm and grows more rapidly with larger graphs.
Moreover, the outcome of this local greedy algorithm depends heavily on the starting point, {as well as the flow and node orders in lines~\ref{alg:hillc:pif} and~\ref{alg:hillc:pin}, respectively.
To improve this suboptimality}, our implementation initially selects $N_{\mathrm{init}}{\,=\,}100$ destination lists at random, evaluates their associated KPIs, and commences the iterations from the best-performing one. 
Nonetheless, a good initialization does not guarantee the achievable performance is good.
Furthermore, we repeat the entire procedure for $N_{\mathrm{rand}}{\,=\,}5$ times and select the best $\bbd$ ever achieved. 
By doing so, we aim to explore various regions of the search space more comprehensively, thereby increasing the likelihood of discovering high-quality solutions.
Based on our empirical observations, while these heuristic tricks can enhance performance, their implementation, even with parallelism, compromises management efficiency.
While we are not delving into this trade-off, it may be of interest for future investigation regarding the balance between achieving higher performance and maintaining efficient management.

\section{Datasets}\label{s:data}

To lay the groundwork for a comprehensive evaluation of GLANCE, it is essential to understand our data.
In the following Section~\ref{s:data:sim}, we introduce the network simulation and data generation pipelines\footnote{Simulation scripts for ns-3 are in C++ and available at \url{https://github.com/bl166/wireless\_digital\_twin\_j}.}.
In Section~\ref{s:data:base}, we explain how we evaluate GLANCE's performance using simulator-based (SB) benchmarks.
In Section~\ref{s:data:base}, configurations used for training GLANCE are discussed. 
By providing details on these key aspects, we aim to facilitate comprehension of data, training, and evaluation procedures.

\subsection{Data generation via simulator}\label{s:data:sim}

\begin{figure}[t]
    \centering
    \begin{subfigure}[t]{\linewidth}
        \centering
        \includegraphics[width=.7\linewidth, trim= 0 2cm 0 1cm]{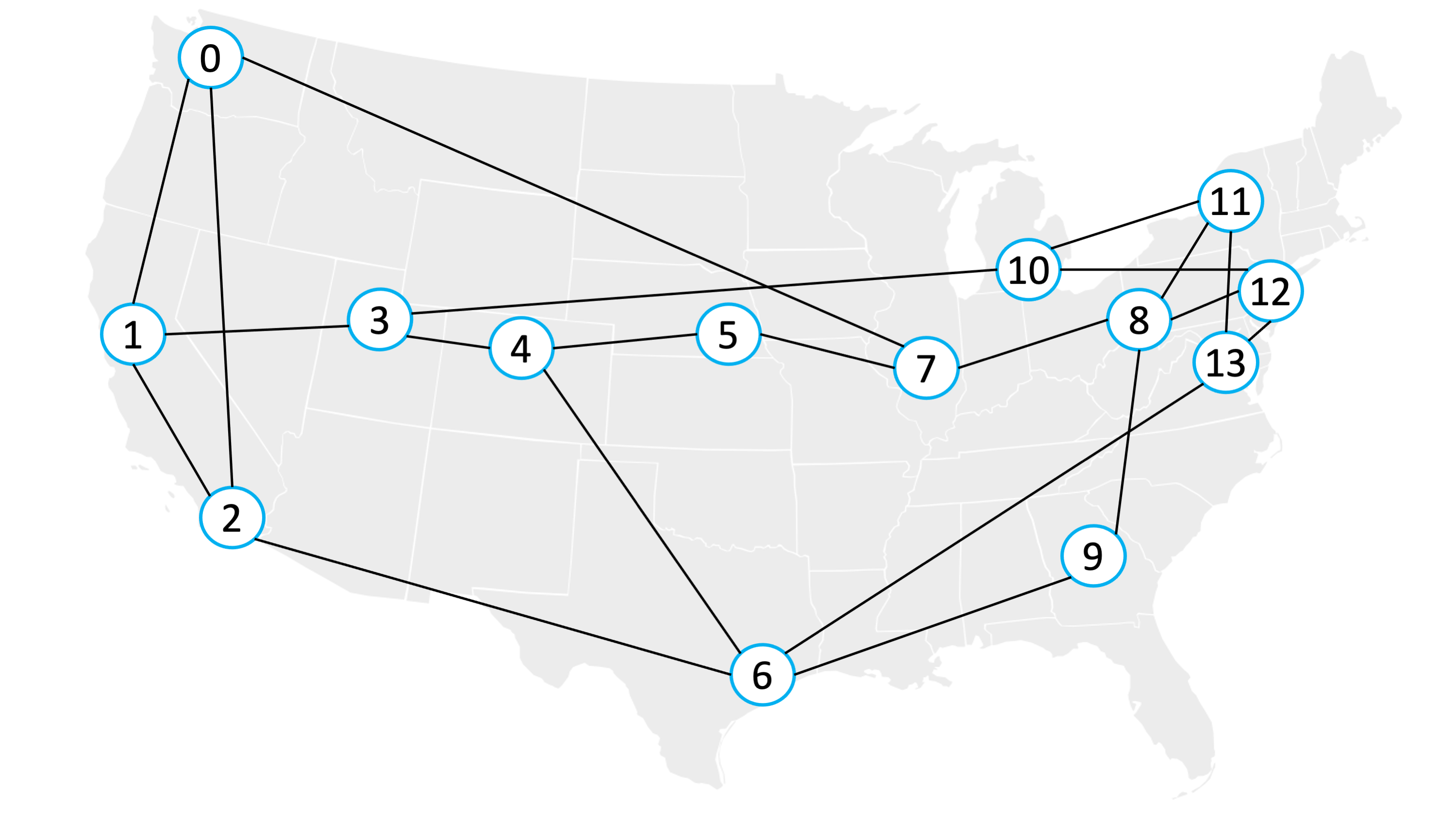}
        \caption{NSFNET with 14 nodes and 42 links (wired).}\label{ff:topo1}
    \end{subfigure}\\
    \begin{subfigure}[t]{\linewidth}
        \centering
        \includegraphics[width=\linewidth, trim= 0 9cm 2cm 1.5cm, clip]{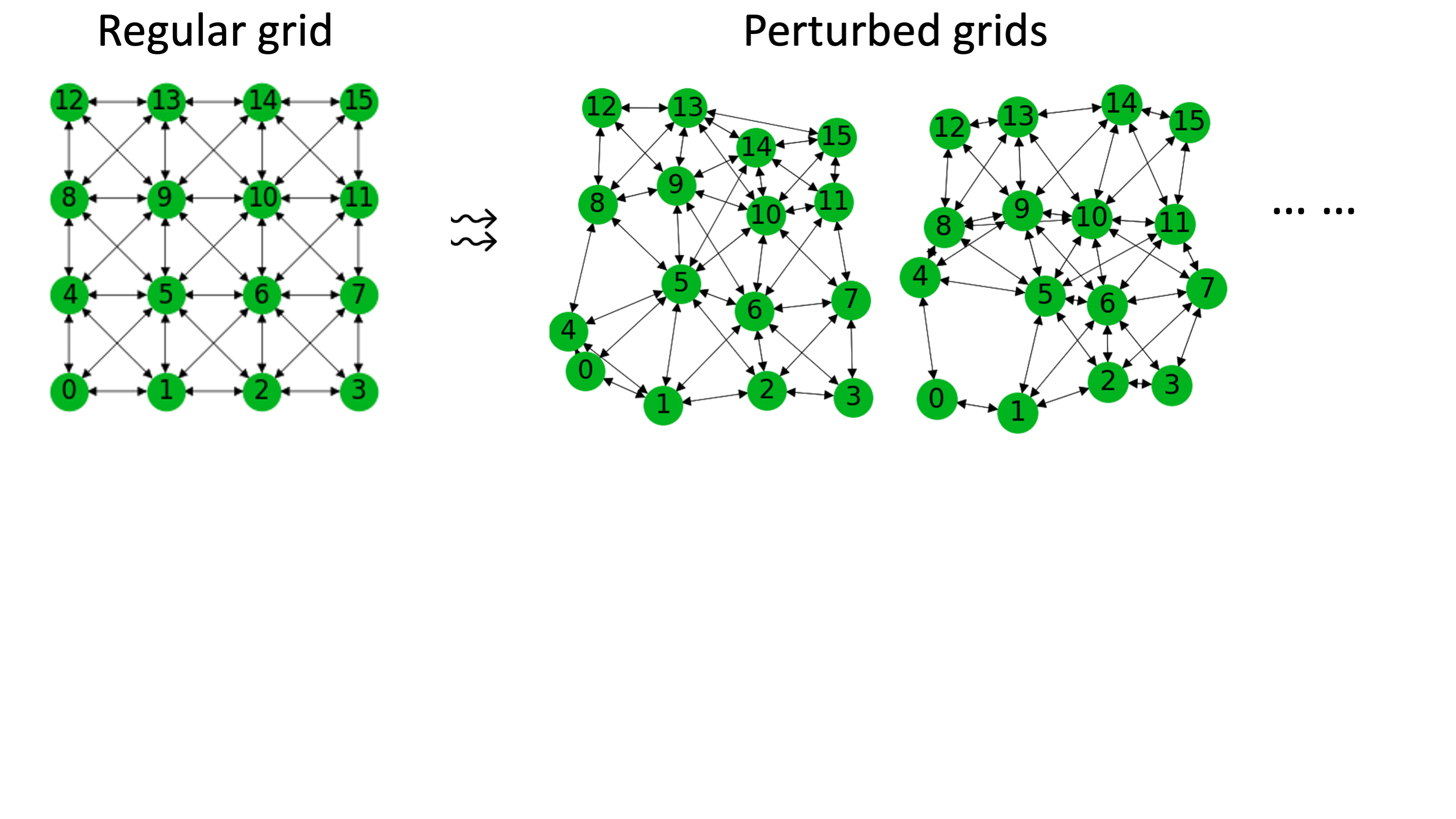}
        \caption{Regular and perturbed grids with 16 nodes (wireless).}\label{ff:topo2}
    \end{subfigure}
    \vspace{-2mm}
    \caption{Topologies for wired and wireless experiments.}
\end{figure}

Our experiments encompass both wired and wireless communication scenarios.
The wired network topology involves the 14-node NSFNET backbone network~\cite{frazer1996nsfnet} in Fig.~\ref{ff:topo1}. 
The more challenging wireless topology pertains to a 16-node grid structure, either regular (RegGrid) or perturbed (PertGrid).
RegGrid has neighboring nodes within the same row or column that are uniformly spaced at a distance of 30 meters.
To get PertGrid topology instances, random perturbations within a 10-meter radius centered at their original positions are introduced, as shown in Fig.~\ref{ff:topo2}.
With RegGrid and PertGrid, we can investigate GLANCE's generalizability to randomness in input entities. 
Wireless path loss follows a log-distance propagation model~\cite{carneiro2010ns}: {$PL{\,=\,}46.67{\,+\,}30\log_{10}d$,} where $d$ is the distance in meters and {$PL$ is the path loss in $\mathrm{dB}$.}
{The ns-3 simulation relies on the path loss. 
While in LDTs,}
as explained in Section~\ref{s:alg}, we model channel information using graphs whose nodes represent mobile devices and edges represent communication links. 
While the wired graph is unweighted, wireless graphs are weighted due to the dominant influence of distance on link strength and network performance.
The weights are defined by the formula {$A_{ij}{\,=\,}1/\log(1+d^2_{ij})$} tailored to ensure numerical stability while preserving an inverse relationship with the path loss.
Both \mbox{ns-3} and GLANCE assume perfect knowledge of {the topology, either via the path loss or the constructed graph.}

In each network instance, $F$ flows are characterized by $F$ unique pairs of source and destination nodes. 
This study specifies $F{\,=\,}10$ uniformly for all samples, though it will be a problem-specific choice in practice.
From source nodes, traffic is generated to their corresponding destinations in an intermittent pattern.
Throughout a simulation, the traffic generators alternate between on and off states.
The duration of each of these states is determined with random variables $t_\text{on}$ and $t_\text{off}$, for which the distribution parameters are predefined. 
During the off state, no traffic is generated. 
During the on state, constant bit rate (CBR) traffic is generated in 210-byte packets at a fixed rate of $100\text{ kb/s}$ (for NSFNET) or $50\text{ kb/s}$ (for grids). 
To be more specific, the on and off times adhere to a compound distribution comprising a continuous exponential distribution and a discrete uniform distribution, which can be denoted as $t_{\{\text{on}, \text{off}\}}\,{\sim}\,\text{Exp}(1/\tau_{\{\text{on}, \text{off}\}})$, where $\tau_{\{\text{on}, \text{off}\}}\,{\sim}\,\ccalU(\{1,10,20\})$.
To clarify further, \mbox{ns-3} and GLANCE assume knowledge on the prerun-sampled $\tau_{\{\text{on}, \text{off}\}}$ values but not the runtime-sampled $t_{\{\text{on}, \text{off}\}}$ values. 
The random sampling of on/off times happens exclusively in \mbox{ns-3} at runtime and is termed {\it traffic resampling}.
{Even if all other topological and flow-level parameters are fixed, this stochastic process contributes variability to KPI outcomes across independent simulation runs.}

The exact routes (or paths) taken are determined by \mbox{ns-3} at runtime, for which we only designate the routing protocol but do not manually enforce any specific routes.
Currently, \mbox{ns-3} employs a modified OLSR protocol.
OLSR proactively discovers the network topology to determine the shortest forwarding paths.
The original OLSR is a dynamic protocol, but we enforce stable routing tables to prevent paths from changing during the same simulation (because GLANCE cannot yet handle dynamic paths).
In essence, the path of a flow should coincide with one -- ns-3 will randomly determine \emph{which} one at runtime -- of the shortest paths identified by any shortest path algorithm such as Dijkstra's.
Thus, if multiple shortest paths exist between a source-destination pair, executing \mbox{ns-3} twice on that instance (with different random seeds) may yield different paths in different runs.
We always monitor the paths determined by \mbox{ns-3} and provide this data to {LDTs, including GLANCE and those against which we compare GLANCE, in order to ensure they use} accurate path information.

\begin{figure}[t]
    \centering
    \includegraphics[width=1\linewidth, trim= 0 5cm 0 .5cm]{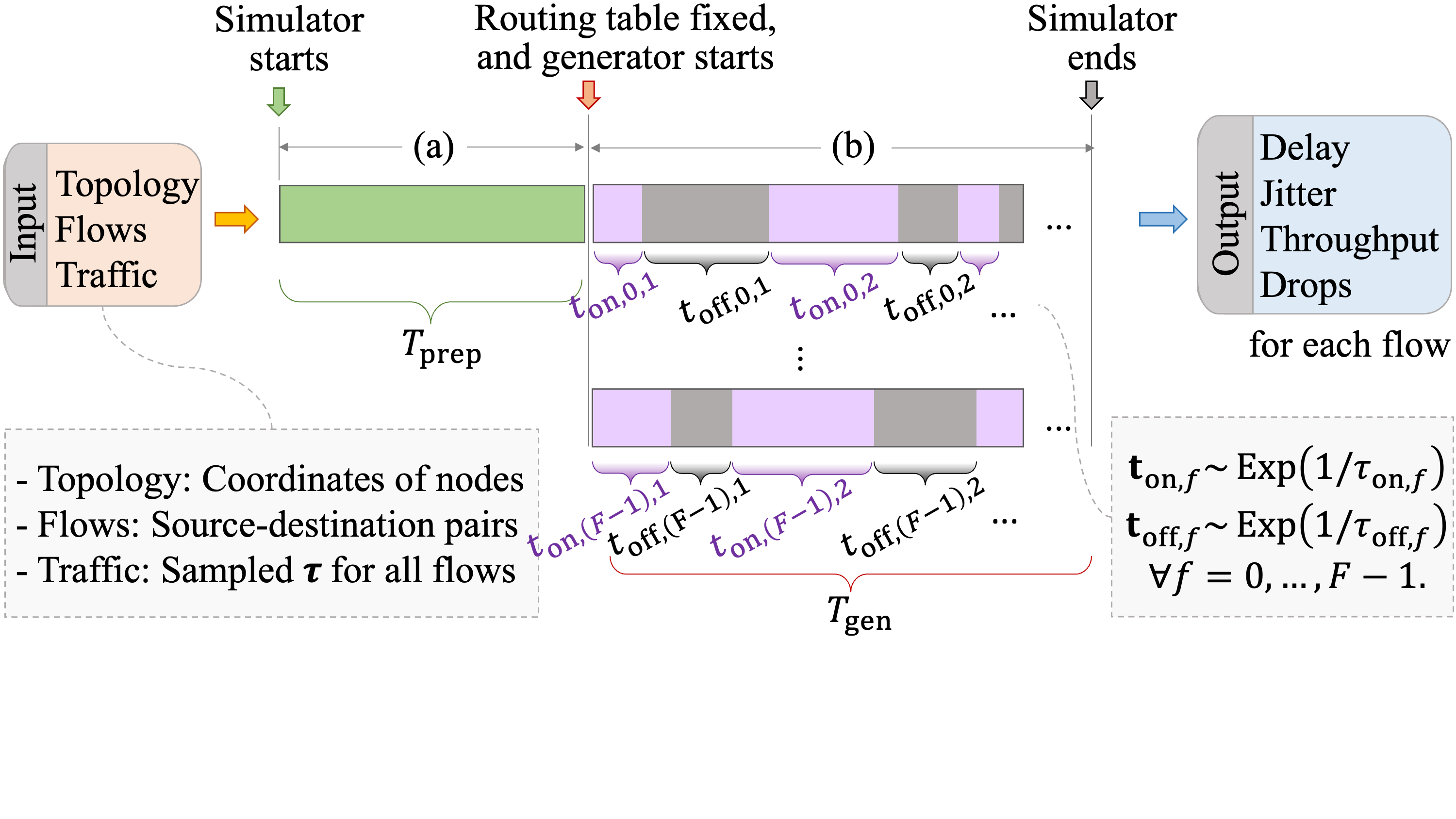}
    \caption{
    Network simulation overview.
    (a) Preparation for $T_\text{prep}{\,=\,}900\text{ simulator-seconds}$, where OLSR discovers the global topology and determines the routing table for this simulation.
    (b) Simulation for $T_\text{gen}{\,=\,}180\text{ simulator-seconds}$, where all source nodes start to generate traffic and forward to their destinations.  
    }\label{f:ns3_sim}
\end{figure}

\begin{figure}[t]
    \centering
    \includegraphics[width=1\linewidth, trim= .25cm .5cm .2cm 0]{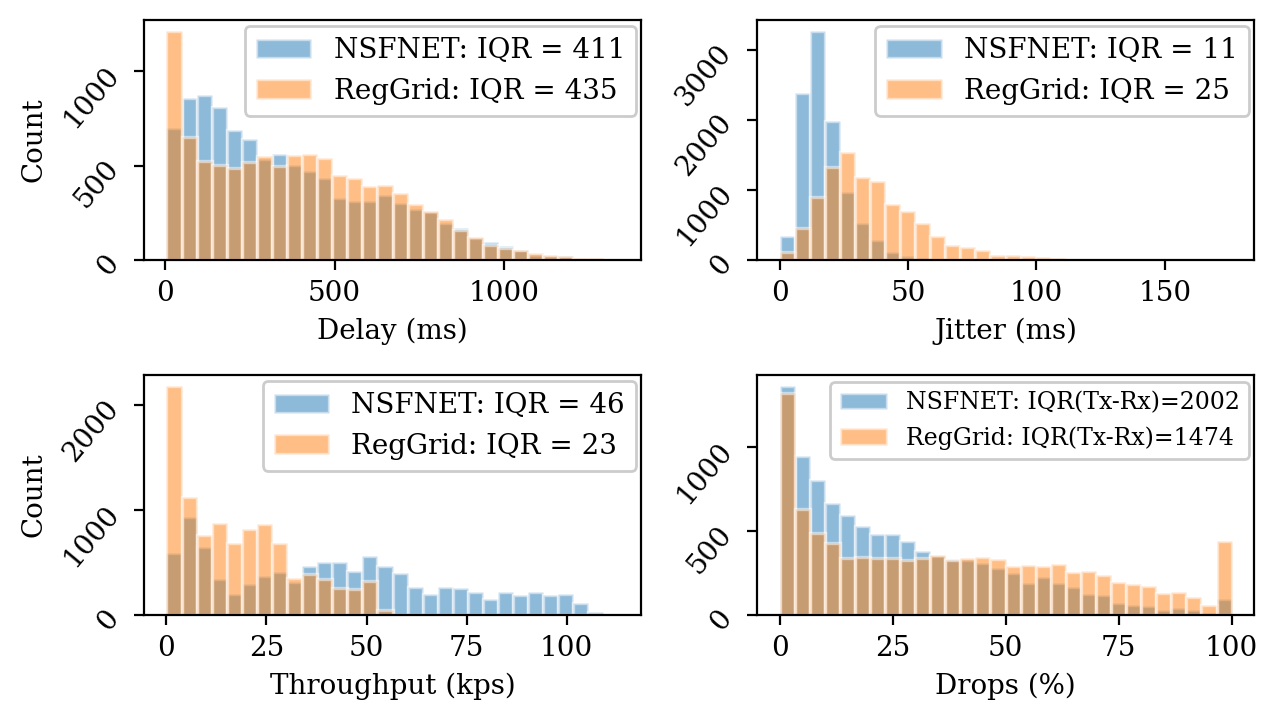}
    \caption{
    Histograms of NSFNET and RegGrid raw KPIs, with IQR annotations.
    For drops, we show IQRs of the absolute numbers of dropped packets, while we plot the distribution of drop rates.
    }\label{f:kpis_dist}
\end{figure}

Now with topology, traffic, and flows (along with a known routing scheme), ns-3 has all the necessary information to conduct simulations over a certain period.
A conceptual simulation process is sketched in Fig.~\ref{f:ns3_sim}.
Following the determination of the routing table, traffic generators in source nodes start to generate traffic intermittently until timeout. 
All flows are monitored for transmitted and received packets (to compute drops), delays, jitter, and throughput (the latter three are calculated as averages over all packets). 
Figure~\ref{f:kpis_dist} displays their {raw value histograms}, taking the fixed-flows-and-topology NSFNET and RegGrid outcomes for example.
These KPIs, alongside the routing table used in this simulation, are saved and pending processing to be used by GLANCE.

{
\remark{
KPIs of different types are likely to vary significantly in scale, as annotated by the interquartile range (IQR) values in Fig.~\ref{f:kpis_dist}.
This necessitates normalization as a pre-processing step for better MTL performance.
To account for outliers, we scale each KPI by its IQR obtained from the training samples.
We compute the mean absolute errors (MAE) between normalized KPIs, termed NMAE, as the major metric of KPI\footnote{{By default, KPI refers to the IQR-normalized KPI in our context of supervision and evaluation.}} evaluation performance. 
}\label{rm:normalize}
}

\subsection{Simulator-based test benchmarks}\label{s:data:base}

The assessment of the twinning performance remains an open challenge. 
Since the goal is to closely emulate \mbox{ns-3} in predicting network KPIs, a straightforward approach would be to compare GLANCE's KPI predictions with \mbox{ns-3} outcomes on the same test samples. 
However, it is acknowledged that GLANCE may not achieve a perfect match due to the stochastic nature of traffic resampling, which reflects real-world randomness. 
Given the inevitable presence of errors, the question arises: how do we ensure that these errors are acceptable or affordable?
In addressing this concern, we need to establish a performance benchmark.
As illustrated in Figure~\ref{f:sb_diag}, this entails additional simulations to mitigate the variance of traffic resampling.
Subsequent steps of this simulator-based (SB) approach are outlined as follows:

\begin{figure}[t]
\begin{subfigure}[b]{\linewidth}
    \centering
    \includegraphics[width=1\linewidth, trim= 0 1cm 0 0]{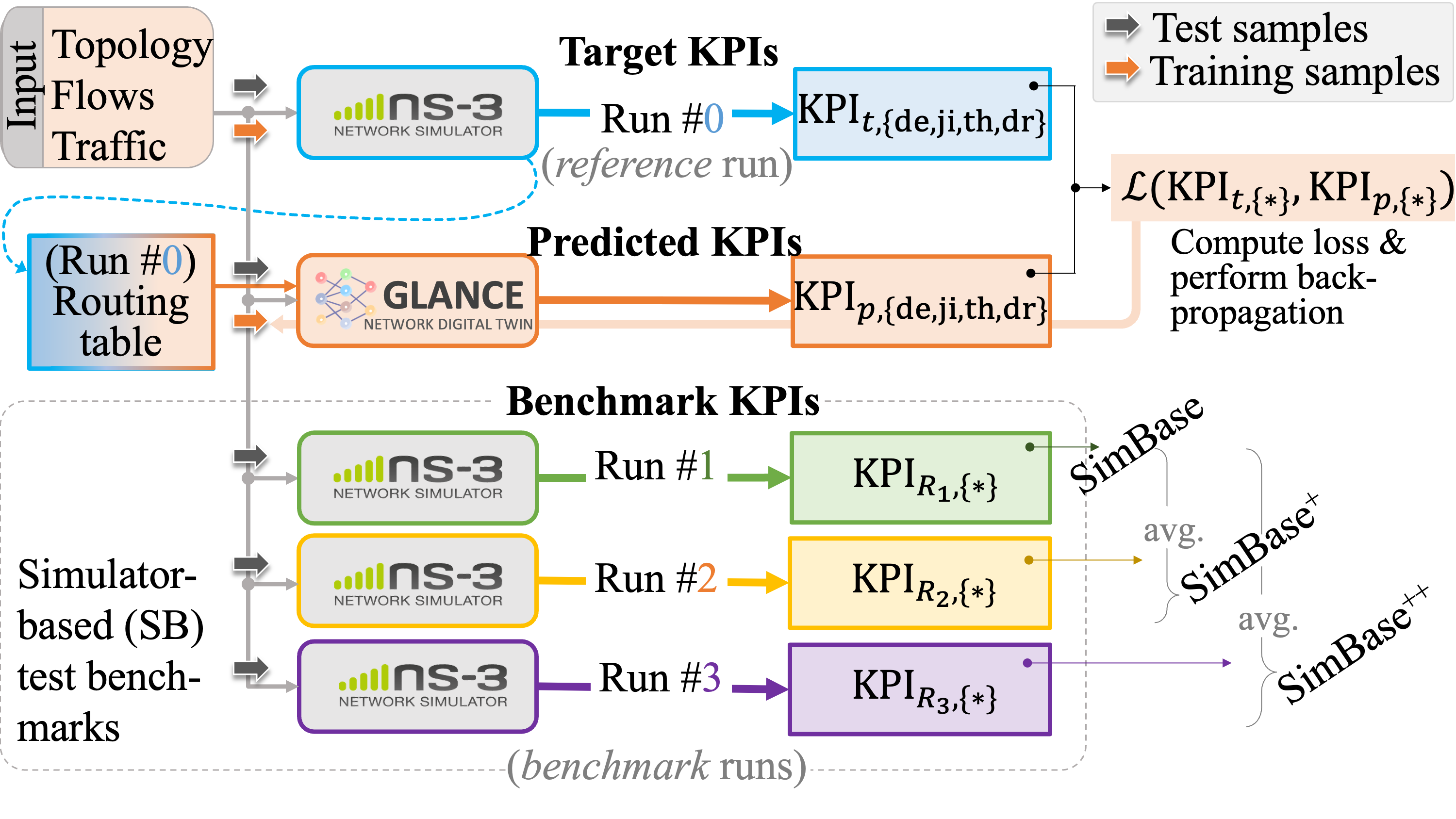}
    \caption{
    Generation of training labels and test benchmarks.
    }\label{f:sb_diag}
    \vspace*{3mm}
\end{subfigure}
\begin{subfigure}[b]{\linewidth}
    \centering
    \includegraphics[width=.95\linewidth, trim= 0 11.3cm 0 0]{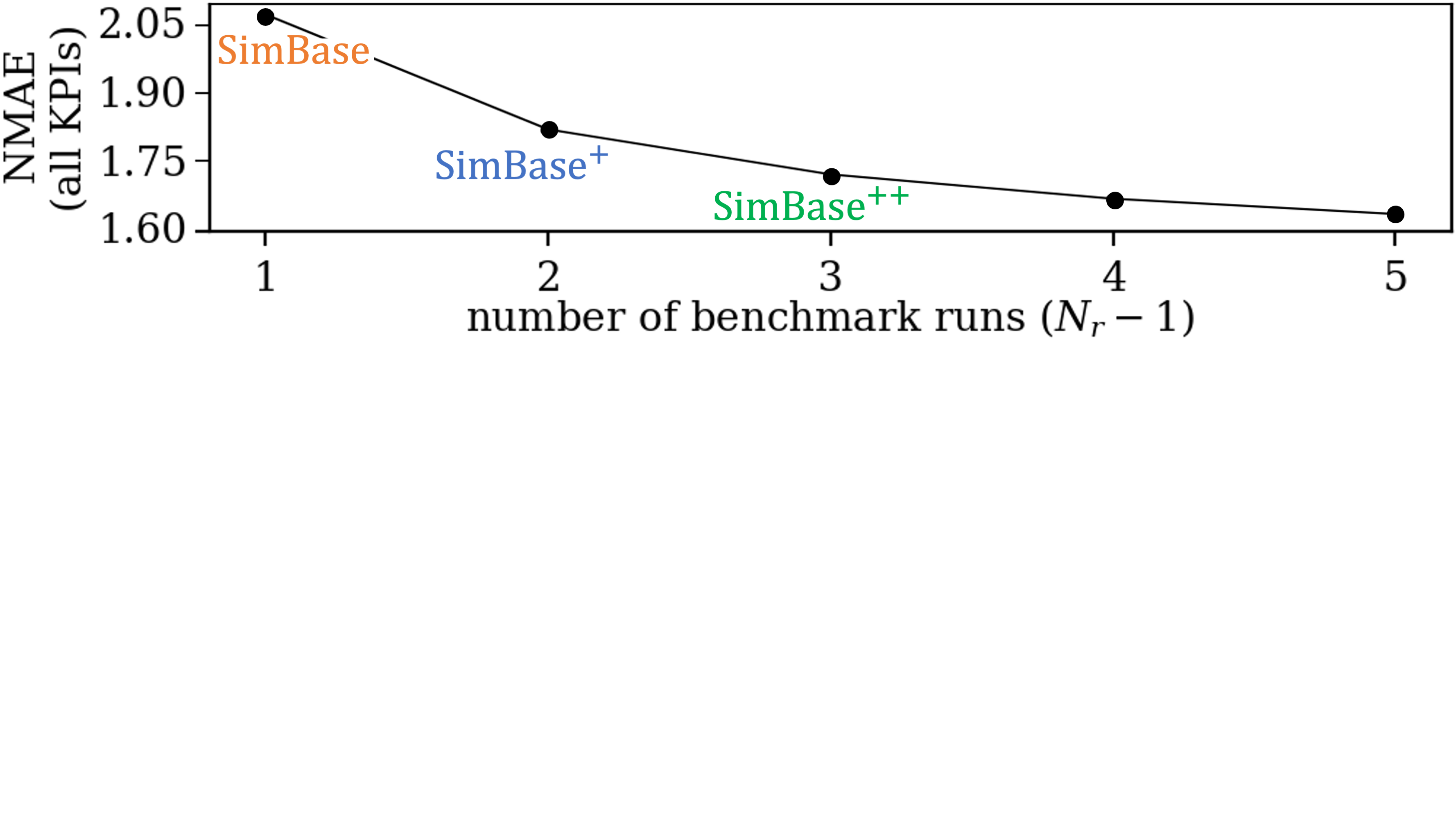}
    \caption{
    NMAE between reference and benchmark KPIs drops as the number of benchmark runs increases. 
    }\label{f:num_runs}
\end{subfigure}
\caption{Training and evaluation with SB benchmarks.} \label{f:sb}
\end{figure}

We provide test samples (test topologies, test flows, and test traffic) to \mbox{ns-3} and execute it for $N_r$ times.
The first run, indexed as $R_0$ or $r{\,=\,}0$, generates target KPIs and is therefore termed as the \emph{reference} run.
Let us denote the target $\KPI_{t,\{*\}}{\,=\,}\KPI_{R_0,\{*\}}$, where the placeholder in the subscript is for specifying the KPI types.
Runs indexed from 1 through $N_r{\,-\,}1$ serve as \emph{benchmark} runs.
Their resulting KPIs can be averaged in subgroups to derive benchmark KPIs with varying degrees of precision (indicated by stacking multiple pluses) depending on the size of the subgroup:
\begin{equation}
    \KPI_{\text{SB,\{de, ji, th, dr}\}}^{\overbrace{+\dots+}^{N_r-2 \text{ pluses}}} = \frac{1}{N_r{-}1}\sum\limits_{r=1}^{N_r-1}\KPI_{R_r,\text{\{de, ji, th, dr}\}}, \forall N_r{\,>\,}1.
\end{equation}
Increasing the number of averaged runs improves the {expected} estimate precision, {given that the expected squared deviation between any two independent and identically distributed (i.i.d.) RVs is greater than the variance of either one of them,}
bringing $\KPI_\text{SB}$ closer to $\KPI_t$.
Figure ~\ref{f:num_runs} exhibits this trend {that brings $\KPI_\text{SB}$ closer to $\KPI_t$} in PertGrid. 
The intuition is to allow stochasticity to average out across multiple runs for better estimation.
In our case, $N_r{\,=\,}4$ for all applicable experiments, resulting in three benchmarks: SimBase, SimBase$^{+}$, and SimBase$^{++}$. 
They correspond to averaging one, two, and three benchmark runs of ns-3 on the test set, respectively. 
Time-consuming may it be, additional benchmark runs should be conducted if higher reliability is required.
As a side note, separate ns-3 runs, though using the same routing protocol, cannot access each other's exact routing tables -- a point mentioned in Section \ref{s:data:sim} but one that may be worth reminding of.
Thus, $R_{\ge1}$ may choose different paths than $R_0$ for the same flow, which could potentially account for a portion of the variance observed in SB outcomes. 

When it comes to the training samples, their traffic data will always differ from that of the test samples. 
The instantaneous network topology and/or flows may remain fixed or vary, depending on the investigation context.
Unlike test data, training data do not necessitate benchmark runs.
As illustrated in Fig.~\ref{f:sb_diag} by the orange arrows symbolizing training samples, their simulations are run only once, and the resulting KPIs are directly used to supervise the training of GLANCE. 
This decision is due to the large number of desired training samples, which makes it uneconomical to run multiple simulations for each. 
In addition, having abundant training samples helps GLANCE learn to handle stochasticity introduced by the simulator.
Instead of aiming for more accurate targets, we focus on providing more diverse data to help GLANCE learn and generalize. 

\subsection{Training configurations}\label{s:data:conf}
\subsubsection{Pre-processing}

Data pre-processing plays a critical role in deep learning, as it does in our study.
It involves a series of steps such as data cleaning, normalization, and augmentation.
Let us revisit the KPI histograms in Fig.~\ref{f:kpis_dist}.
In these distributions, we have excluded outlier samples (less than $1\%$) with maximum delay greater than $2000\mathrm{ ms}$ or jitter greater than $200\mathrm{ ms}$.
Besides, concurrency issues, simulation uncertainty, resource contentions, and unexpected simulator failures may result in missing flows, affecting less than $1\text{\textperthousand}$ simulations. 
Now, we must consider how to address the outlying or missing flows in samples.
If it is a training sample, we discard it. 
However, before discarding any test sample that has undergone parallel benchmark runs, we attempt to impute it using averaged results from other benchmark runs that are valid for the same input.
Note that the imputation does not involve the reference run, in order to avoid information leaks and ensure fair comparisons. 
Along with the normalization step introduced in Remark~\ref{rm:normalize}, we have balanced the discrepant KPI scales that could hinder the embedding layers' ability to generalize in MTL.

\subsubsection{Hyperparameters}

Table~\ref{tab:hyperparams} lists important hyperparameters for each dataset utilized in Sections~\ref{s:exp:emb} through \ref{s:exp:gen}.
{If the topology or flows of a dataset is `fixed', all samples share the same data for the corresponding entry.
If marked as `random', each sample has independently sampled data for that entry.} 
This table also states dataset sizes, model layer dimensions, regularization coefficients, and learning rates. 
Overall, GLANCE contains $T{\,=\,}3$ embedding layers and is trained with a batch size of 10 over 100 epochs for 4-fold cross-validation (CV). 
The longest path present in all datasets has 3 in-path links, setting a cap on the longest path that GLANCE can infer. 
As an additional layer of regularization, we adopt parameter sharing to combat overparameterization. 
This ensures that $\text{RNN}_t$, $\text{MLP}_t$, and $\text{GCN}_t$ remain identical for all $t$, effectively reducing the number of embedding parameters by a factor of $T$.

\begin{table*}[t]
    \caption{
    Summary of experiment data and corresponding hyperparameters for training GLANCE.
    }\label{tab:hyperparams}
    \vspace{-.5em}
    \centering
    \small
    \resizebox{\linewidth}{!}{%
        \begin{tabular}{c|c|c|c|ccc|c|c|cc|c}
        \hline
             \multirow{2}{*}{Section} & \multirow{2}{*}{Topology} &  \multirow{2}{*}{Flows} & \# of Samples &  \multicolumn{3}{c|}{Emb. Dimensions} &  \multicolumn{2}{c|}{MLP Sizes} & \multicolumn{2}{c|}{L2 Regularization} & Learning \\
        \cline{5-11}
         & &  &  (Test\,/\,Val.\,/\,Train) &   Node & Link & Path & Link & Readout & Link & Readout &  Rate  \\
         \hline
        \multirow{2}{*}{\ref{s:exp:emb},\ref{s:exp:comp}} & NSFNET (Fixed)& \multirow{2}{*}{Fixed} & \multirow{2}{*}{2.5K (0.5\,/\,0.5\,/\,1.5 K)} & \multirow{2}{*}{16} & \multirow{2}{*}{16} & \multirow{2}{*}{32} & \multirow{4}{*}{ [64,32,16]} & \multirow{2}{*}{[32, 64, 128, 32]} &\multirow{2}{*}{$10^{-3}$}& \multirow{2}{*}{$10^{-4}$} & $10^{-3}$ \\
        \cline{2-2}\cline{12-12}
         & \multirow{2}{*}{RegGrid (Fixed)} &      &      &    &    &    &    & & &  &  \multirow{3}{*}{$5{\times}10^{-4}$} \\
        \cline{1-1}\cline{3-7}\cline{9-11}
        \ref{s:exp:gen:flow}&        & Random & \multirow{2}{*}{5K (1\,/\,1\,/\,3 K)} & \multirow{2}{*}{32} & \multirow{2}{*}{32} & \multirow{2}{*}{64} &  &\multirow{2}{*}{[64, 128, 128, 32]}& \multirow{2}{*}{$10^{-4}$}& \multirow{2}{*}{$10^{-5}$}  &    \\
         \cline{1-3}
        \ref{s:exp:gen:topo}& PertGrid (Random) & Fixed & & & & &  & & &  &    \\
        \hline
        \end{tabular}
    }
\vspace*{-1em}
\end{table*}

\section{Numerical experiments}\label{s:exp}

We conduct a thorough evaluation of GLANCE to explore its multifaceted capabilities. 
In Section~\ref{s:exp:emb}, we showcase that GLANCE adeptly learns predictive network embeddings jointly from and for multiple KPIs, either directly or through TL. 
Section ~\ref{s:exp:comp} highlights GLANCE's superior performance via comparisons with {other DTs}. 
In Section~\ref{s:exp:gen}, we demonstrate GLANCE's generalizability when applied to random flows and topologies.
In Section~\ref{s:exp:app} presents compelling evidence of GLANCE's effectiveness in two network management applications. 
In particular, we employ it to control the traffic loads and destinations, aiming to achieve a target KPI profile in each case.

\subsection{Learning universal embeddings}\label{s:exp:emb}

\begin{figure}[t]
    \centering
    \begin{subfigure}[t]{0.48\linewidth}
        \centering
        \includegraphics[height=.8\linewidth, trim= .25cm .2cm .2cm 0, clip]{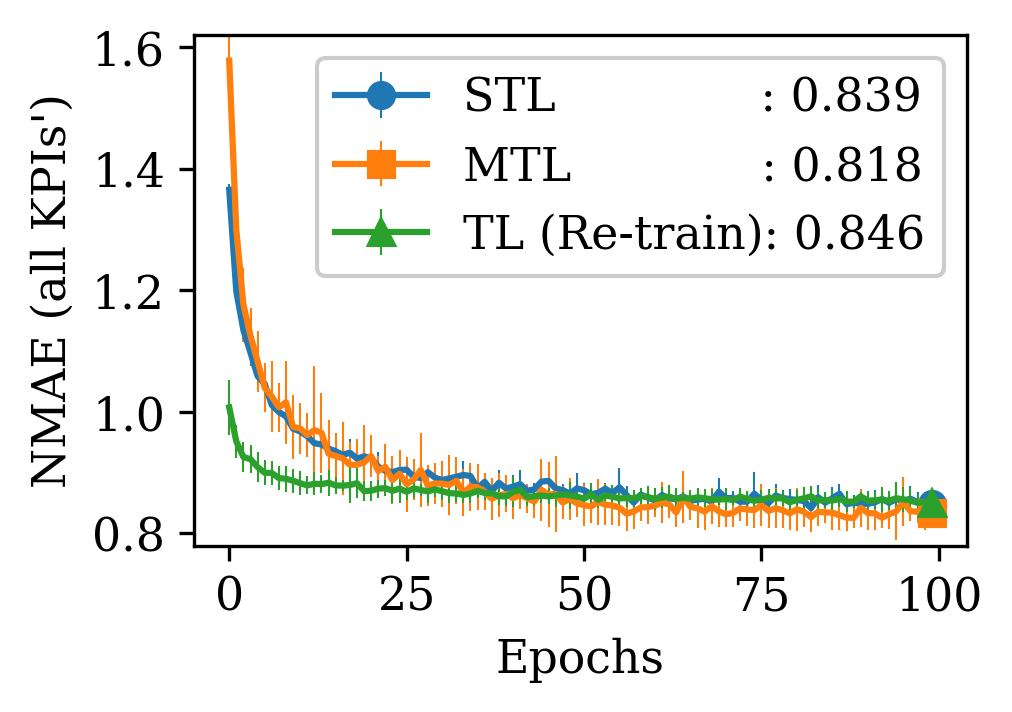}
        \caption{NSFNET.}\label{f:learning_strategy_a}
    \end{subfigure}%
    ~
    \begin{subfigure}[t]{0.48\linewidth}
        \centering\hfill
        \includegraphics[height=.8\linewidth, trim= 1.55cm .2cm .2cm 0, clip]{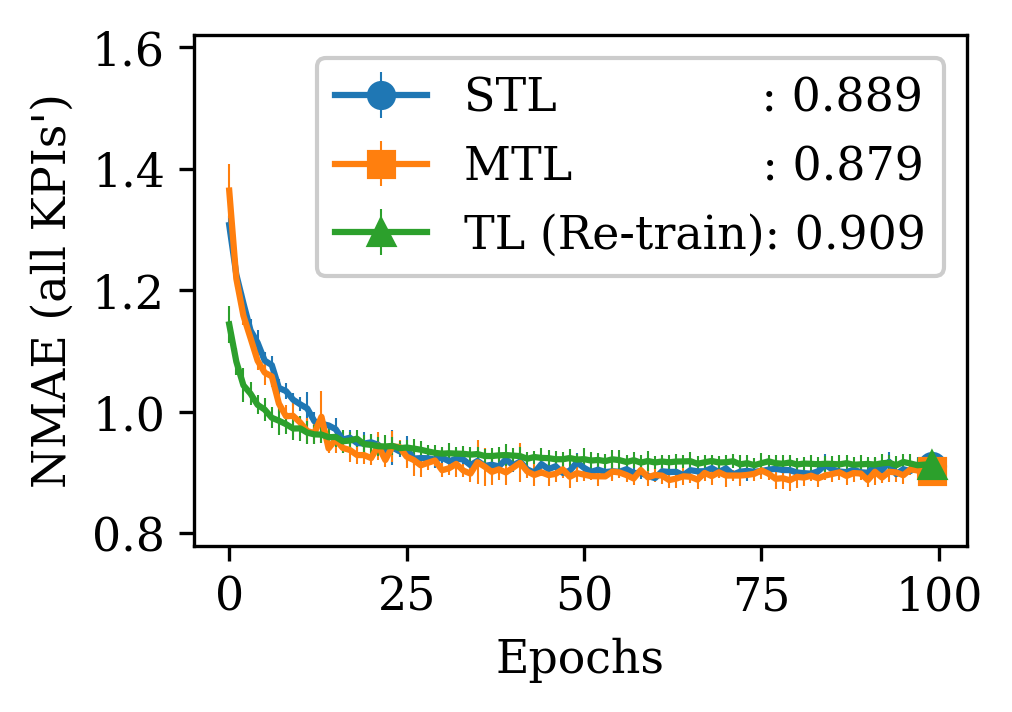}
        \caption{RegGrid.}\label{f:learning_strategy_b}
    \end{subfigure}
    \vspace{-.5em}
    \caption{Learning strategies compared considering the total NMAEs across all KPI types. 
    Individual KPIs are consistent with this trend. }\label{f:learning_strategy}
\end{figure}

The {\it universality} of learned network embeddings emphasizes their utility in predicting not only the trained KPI types but also the unseen ones via TL.
Concerning the input entities, we fix the topology and flows to narrow GLANCE's focus down for a pinpointed investigation into the impact of different learning strategies. 
We present a comparison of prediction performance across all KPI types, as indicated by their summed NMAEs, for the three training strategies discussed in Section~\ref{ss:learning_strategies}.
For TL, the pre-training curves are omitted, and we show the re-training phase only. 
In Fig.~\ref{f:learning_strategy}, we plot for both NSFNET and RegGrid topologies the validation NMAEs against the number of completed epochs.
We observe the lowest achieved NMAE value in each CV fold, calculate their average, and annotate it on the plots.

The following observations generally hold true for both NSFNET and RegGrid.
First, the MTL performance ends up statistically better than or equal to that of STL.
Secondly, TL re-training starts off at a better position than STL and MTL, converges more rapidly in the early epochs, and eventually aligns with at least STL.
{These results align with many desirable properties of MTL and TL that have been observed in other domains, such as~\cite{oh2022improving,lamsiyah2023unsupervised}.
In our case, it suggests} that GLANCE is adept at comprehending the fundamental network structure and features.
It can derive universal embeddings that portray the inherent network characteristics, rather than being overly focused on specific KPIs of supervision.
In summary, when sufficient training data are available for both the target and other KPIs, MTL generally achieves the best and most efficient learning performance.
In the lack of target KPI labels, it is still possible to fulfill a slightly compromised performance via TL.

\subsection{Comparing to baseline and benchmark results}\label{s:exp:comp}

The previous section has shown that joint multi-tasking is the preferred training strategy for GLANCE.
Next, we highlight the effectiveness of GLANCE's specialized architecture.
Enhanced with graph learning components for handling node interaction and interference, this architecture can incorporate {fundamental domain knowledge that helps achieve more accurate multi-KPI predictions}.
In addition to the SB benchmarks introduced in Section~\ref{s:data:base} (SimBase, SimBase$^{+}$, and SimBase$^{++}$), we include the following LDTs as supplementary performance baselines.

\begin{enumerate}[wide,label={\arabic*)},labelindent=0pt]
    \item \textbf{RouteNet}: 
    An architecture similar to GLANCE, sharing identical input and output representations, but that ignores node embeddings.
    In other words, removing Step~(iii) and $\bbh_n^t$ from Algorithm~\eqref{alg:cap} and adjusting the impacted dimensions will result in this algorithm. 
    Its ability to capture relational link-path information is more suitable for wired settings in which interference between nodes in different paths is less of a concern. 
    {For more details on RouteNet, please refer to~\cite{rusek2020routenet}.}

    \vspace*{1mm}
    \item \textbf{RouteNet-F}: 
    An extension and the latest follow-up work of RouteNet, of which the intuition falls between RouteNet and GLANCE.
    Similar to GLANCE, it has explicit node embeddings in addition to path and link embeddings.
    Unlike GLANCE, it does not utilize the full network topology. 
    Instead, it only uses path information, similar to RouteNet.
    Additionally, its link and node subnets both take the form of GRU cells. 
    Moreover, rather than directly reading out path embeddings as a whole, RouteNet-F's readout blocks operate on link embeddings and produce KPI predictions in an additive manner, {which are potentially more versatile than those in RouteNet and GLANCE.
    This additive design} aligns with the nature of some KPIs, where, for example, the delay of a flow is the sum of delays across all links in the corresponding path.
    For more details on RouteNet-F, please refer to~\cite{ferriol2023routenet}.

    \vspace*{1mm}
    \item \textbf{GNN}:
    A simple graph neural network (GNN) in multi-task mode. 
    Three 96-channel graph convolutional layers are followed by $K$ parallel dense layers each reading one KPI type out.
    Its input, or the node features denoted as $\bbx^{(j)}$ for node $n_j$, consists of $2F$-dimensional vectors incorporating traffic data:
    \begin{equation}\label{e:gnn_node_features}
        x^{(j)}_{\{2f,2f{+}1\}}=
        \begin{cases}
          \tau^{(f)}_\onoff, & \text{if } n_j\in p_f,\\
          0, & \text{if } n_j\notin p_f, \forall\,f=0,...,F-1.
        \end{cases}
    \end{equation}
    With this input formulation, the need for a path length limit dissipates. 
    However, the fixed order of flows in~\eqref{e:gnn_node_features} renders it non-equivariant to permutations.
    Another flaw resides in its predefined input dimension. 
    This limitation curbs its usefulness considering the common demand for flexibility regarding the number of flows in practice.
    Table~\ref{tab:lb-comp} summarizes key differences between {the candidate LDTs}.
\end{enumerate}

\begin{table}[H]
\vspace{-.5em}
    \caption{
    Comparison of LDTs. 
    The capacity attribute is quantified by the number of trainable parameters. 
    {The GLANCE capacity shown in this table is with concern to Section~\ref{s:exp:comp}, which corresponds to the compact configuration in Table~\ref{tab:hyperparams}}. 
    }\label{tab:lb-comp}
    \vspace{-.5em}
    \centering
    \small
    \resizebox{\linewidth}{!}{%
        \begin{tabular}{l|c|c|c|c}
        \hline
             {\diagbox[width=2cm]{Method}{Feature}} &  Specialization & Node emb. & Full topology & Capacity \\
        \hline
        GLANCE & \checkmark & \checkmark & \checkmark & \num{43108} \\
        RouteNet & \checkmark & -- & -- & \num{40532} \\
        RouteNet-F & \checkmark & \checkmark & -- & \num{29524}\\
        GNN & -- & \checkmark & \checkmark & \num{24520} \\
        \hline
        \end{tabular}
    }
\vspace{-.5em}
\end{table}

\begin{figure*}[t]
\centering
    \begin{tikzonimage}[
        width=\linewidth, 
        trim=0cm 0 0 .5cm
    ]{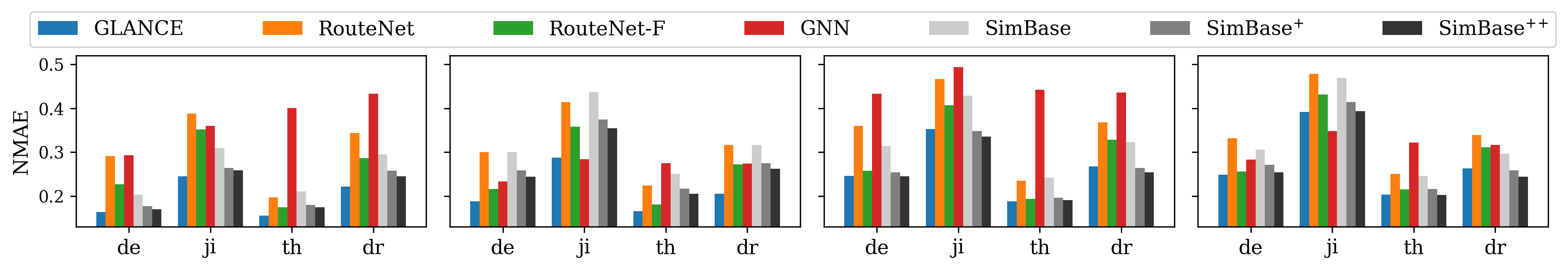}
    \node at (0.15, -0.03) {(a)~NSFNET};
    \node at (0.40, -0.03) {(b)~RegGrid};
    \node at (0.64, -0.03) {(c)~RegGrid (random-flows)};
    \node at (0.87, -0.03) {(d)~PertGrid};
    \end{tikzonimage}
    \vspace*{-2em}
    \caption{
    Bar-plot visualizations corresponding to the NMAE mean values for different datasets.
    }
    \label{f:tabfig}
\vspace*{-1em}
\end{figure*}
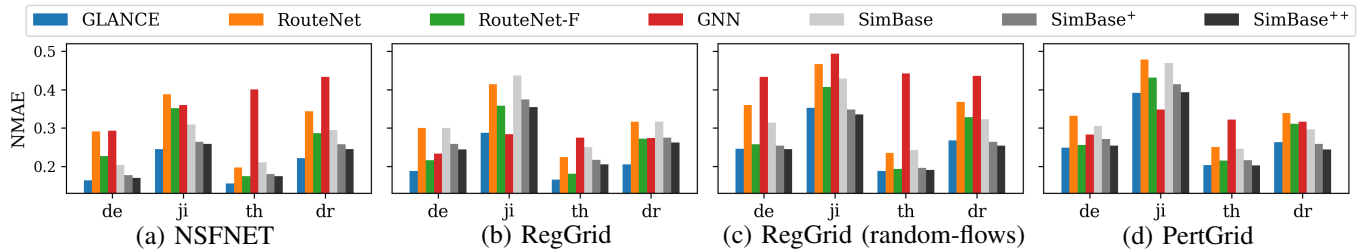

\subsubsection{Wired network}\label{s:exp:comp:nsfnet}
To present Fig.~\ref{f:tabfig}a, we configure a wired network scenario identical to that used in Fig.~\ref{f:learning_strategy_a}.
A greater gap that an LDT gets over SB benchmarks for a KPI indicates better performance that the twin can achieve for that KPI.
Taking throughput for example, GLANCE attains an NMAE of 0.156 which is not only lower than that of either RouteNet(-F) or GNN but also $10.9\%$ lower than SimBase$^{++}$'s 0.175.
For other KPIs, the relative performance of GLANCE to SB is consistent, suggesting that {GLANCE can estimate NSFNET KPIs more accurately than running ns-3 three times.}

{As for the competing LDTs, their advantage over SB methods is mostly observed in the inference efficiency.}
While training LDTs may take hours, once deployed, they can complete predictions in tens of milliseconds. 
In contrast, SimBase requires hundreds of seconds, not to mention SimBase$^{+}$ and SimBase$^{++}$ which double and triple the time taken by SimBase, respectively.
In this case, we have verified with six benchmark runs that SimBase$^{5+}$ can eventually approach the performance of GLANCE.
{Intuitively speaking, this is because GLANCE learned the mean from training across many instances of ns-3.}
Nevertheless, resorting to SimBase$^{5+}$ would be impractical due to its time-consuming nature.
Intuitively, LDTs trade-off inference time for training time, where the latter can be conducted offline with less constraints on computational resources and time.

\subsubsection{Wireless network}\label{s:exp:comp:wifi}

For Fig.~\ref{f:tabfig}b, we perform the same training and benchmarking procedures in the wireless RegGrid network.
Alongside GLANCE's strengthened lead across all KPI types, RouteNet exhibits overall advantages over SimBase, and RouteNet-F is comparable to SimBase$^{++}$, as well. 
The relatively greater degradation in SB performance is likely attributed to the increased indeterminacy in the routing paths.
In this highly regular topology, simulated link strengths are essentially binary: either a stronger link between adjacent nodes in the same row/column or a weaker one between nodes on the grid units' diagonals.
Hence, there is a higher chance that multiple shortest paths may exist for some flows, leading to a greater likelihood of ns-3 benchmark runs taking different routes than the reference run despite identical source and destination nodes (Fig.~\ref{f:sb_diag}).
While the topological regularity is adverse to SB performance, it does not affect LDTs as much, since the latter has access to the reference routing table. 
Leveraging that information, LDTs can aggregate the accurate link embeddings into path embeddings.
Therefore, they may gain extra benefits over SB here than in the NSFNET case where the routing was less indeterministic.\looseness=-1

To sum up, GLANCE takes a remarkable lead in performance as an LDT to ns-3, owing to its specialized architecture and utilization of the underlying network topology.
Given unlimited time or resources, SimBase$^{n+}$ with a sufficiently large $n$ may be a more accurate option.
In practice, however, many applications of network evaluators demand near real-time processing speed, as will be shown by two examples in Section~\ref{s:exp:app}. 
In such cases, the potential accuracy advantage offered by SB benchmarks is often overshadowed by the efficiency advantage of LDTs.

\subsection{Generalizing to random inputs}\label{s:exp:gen}

In the following, we consider more realistic scenarios characterized by a higher level of stochasticity in the inputs, where testing flows or topologies are unseen during training. 
To bolster generalizability, we {randomly sample different flows and topologies} 
and include more data samples in the training set.
Hyperparameters are also tuned to increase GLANCE's capacity for the expanded input space, as given in Table~\ref{tab:hyperparams}. 
Respectively, we conduct two experiments to explore randomness in flows and in topologies within wireless networks.

\subsubsection{Random flows}\label{s:exp:gen:flow}

Previously, all instances featured an identical set of 10 flows, which is a mere experimental setting that can be convincingly argued as unrealistic. 
In addressing this concern, we now randomize both the source and destination nodes of the flows while ensuring their uniqueness within an instance.
The underlying topology remains to be the fixed RegGrid. 
\mbox{Figure~\ref{f:tabfig}c} compares {DT performance.
Between LB and SB approaches, SB generally observes} a notable increase in the relative-to-LB performance (cf. Section~\ref{s:exp:comp}). 
{SimBase$^{++}$ prevails in all KPIs but throughput.}
Following closely are GLANCE and SimBase$^{+}$, which show roughly competitive performance with each other.
What used to perform relatively well in the fixed-flows case, namely GNN, falls behind in this scenario due to its lack of flexibility in incorporating the constantly changing link sequences.

Comparing these results with those of Fig.~\ref{f:tabfig}b, drawing a clear line between {LB and SB twinning} is whether their performance is negatively influenced by random flows.
All LDTs exhibit {higher errors} with random flows, while SB benchmarks demonstrate different or contrasting behaviors.
Viewed from a stochastic-learning perspective, the flow input is now uniformly sampled from the vast space of $\binom{N(N-1)}{F}$ possible combinations, where $N{\,=\,}16$ nodes and $F{\,=\,}10$ flows. 
This entails a very large sampling space compared to the scale of our training set, posing a significant challenge to the generalizability power of LDTs.
On the contrary, SB methods are not affected in a deterministic manner as LDTs are. 
To explain the improvement observed in SB performance, it may be because the previous fixed flows were initially a hard case resulting in higher simulation variability than average.
Consequently, the introduction of other flows may lead to less variability, thereby reducing NMAEs overall. 
However, if the fixed flows were an easy set, the direction of performance change could be reversed.
In summary, random flows are likely to adversely affect LDTs more than SB benchmarks.

\subsubsection{Random topologies}\label{s:exp:gen:topo}

In the following, we fix the flows and allow LDTs to concentrate on generalizing against random topologies. 
We introduce random positional perturbations to each node in the underlying RegGrid topology, referred to as PertGrid, as described in Section~\ref{s:data:sim} and depicted in Fig.~\ref{ff:topo2}.
\mbox{Figure~\ref{f:tabfig}d} presents the new NMAEs under this setting.
In general, the LB-versus-SB trend is similar to that in Section~\ref{s:exp:gen:flow}: a universal performance degradation for LDTs.
With that, GLANCE now leads only in delay and is tied with SimBase$^{++}$ in throughput.
Likewise, RouteNet outperformed SimBase in one or more KPIs in RegGrid (Section~\ref{s:exp:comp:wifi}) but is completely falling behind the latter in PertGrid.
RouteNet-F consistently outperforms RouteNet and underperforms GLANCE.
GNN generalizes much better to random topologies than to random flows, which comes as no surprise.

To summarize this section, the observations above are evidence of the inherent difficulty in predicting network KPIs from random inputs. 
Despite this challenge, our proposed method stands out as the most generalizable and comprehensive LDT. 
Architecture-wise, RouteNet(-F) specializes less in topology, whereas GNN tends not to focus on flows.
In contrast, GLANCE is expressive in both flows and topology, thus meriting a powerful strength to mirror the performance of SB benchmarks while significantly reducing time costs.
In the next and final set of experiments, we will harness the predictive power of GLANCE by integrating it into an optimization framework for managing communication networks, aiming to achieve a target profile on KPIs.

\subsection{Applications in network management}\label{s:exp:app}

\begin{figure}[t]
    \centering
    \includegraphics[width=1\linewidth, trim= 0 7.5cm 0 .3cm]{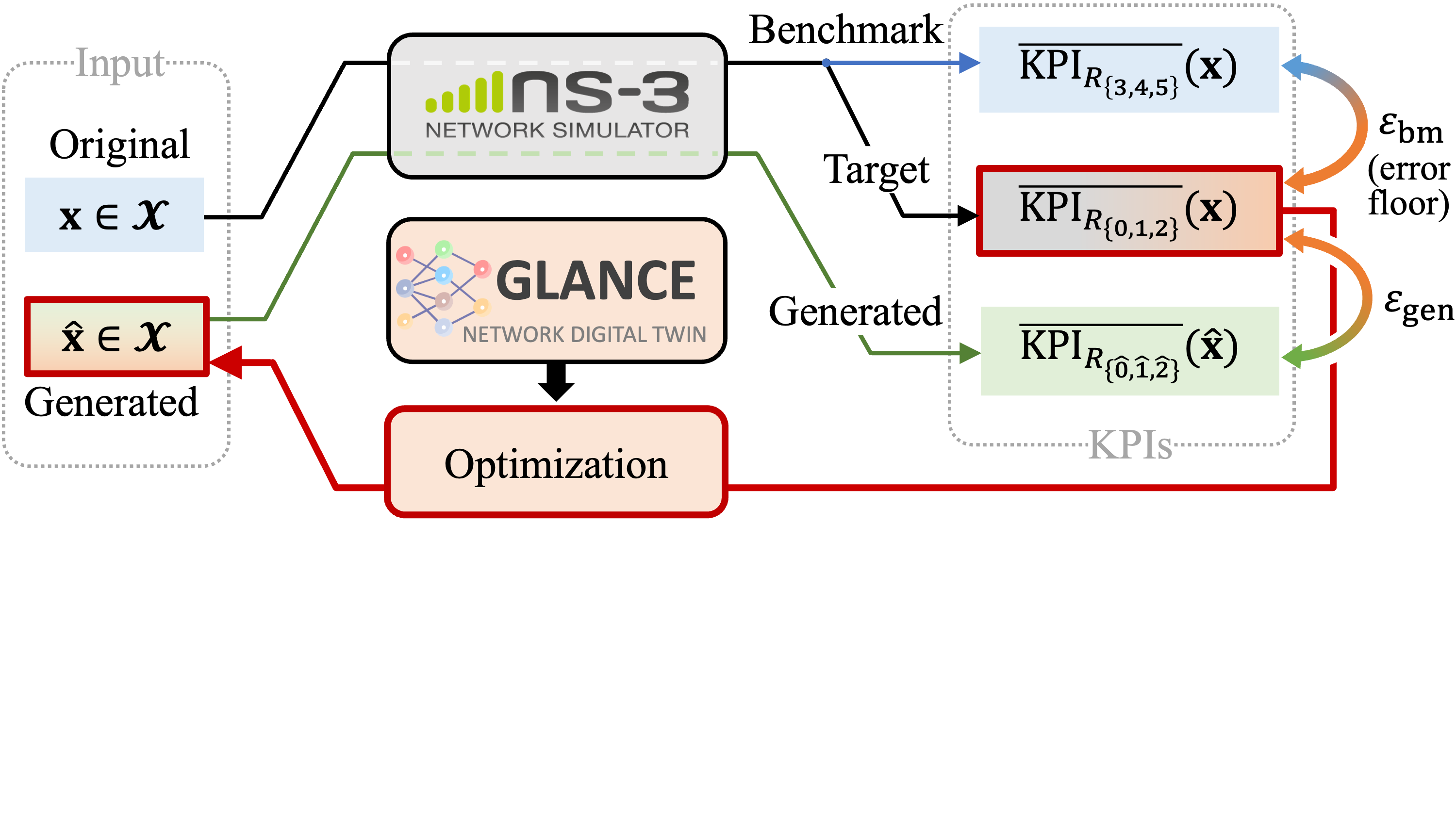}
    \caption{
    Diagram of network management using GLANCE, including both optimation and evaluation regimes.
    }\label{f:manage_diag}
\end{figure}

Backed by the promising results in Sections~\ref{s:exp:emb} through~\ref{s:exp:gen}, we now exploit GLANCE's predictive ability to manage the loads and destinations of flows, respectively, formulated within an optimization framework (Section~\ref{ss:manage}). 
With the goal of finding inputs whose resulting KPIs through the network system of interest closely match some target values, the first consideration is to set feasible KPI targets. 
In real-world scenarios, engineers responsible for system maintenance may leverage domain knowledge to propose desired and attainable KPI targets.
In our research context, we establish a target KPI profile by consulting ns-3.
For an instance depicted in Fig.~\ref{f:manage_diag}, we start with an original input $\bbx$ (either traffic $\bbtau{\,\in\,}\mbR^{2F}$ or flow $\bbf_\dst{\,\in\,}\ccalN^{F}$) for which the 3-run-averaged KPIs are obtained by ns-3 and will serve as the target KPIs, denoted as $$\bbk_\targ{\,=\,}\overline{\KPI_{R_{\{0,1,2\}}}}(\bbx),$$ whose dimension depends on the number of KPI types considered herein.
After the optimization completes, the generated input is denoted as $\hbx$, of which the evaluation also relies on ns-3 such that $$\hbk_\gen{\,=\,}\overline{\KPI_{R_{\{\hat{0},\hat{1},\hat{2}\}}}}(\hbx).$$
Let us define a generation error $\varepsilon_\gen{\,=\,}MAE(\hbk_\gen, \bbk_\targ).$
Additionally, we generate a KPI-performance benchmark in 
another 3-run averaging process based on the original input $\bbx$, namely $$\bbk_\mathrm{bm}{\,=\,}\overline{\KPI_{R_{\{3,4,5\}}}}(\bbx),$$ thus obtaining a benchmark error $\varepsilon_\mathrm{bm}{\,=\,}MAE(\bbk_\mathrm{bm},\bbk_\targ).$
The proximity of $\hbk_\gen$ to $\bbk_\targ$, or comparisons between $\varepsilon_\gen$ and $\varepsilon_\mathrm{bm}$, can offer insights into the management performance.
The specific optimization and evaluation operations depicted in this diagram depend on the specific content being optimized and the algorithm employed.\looseness=-1

\subsubsection{Traffic load optimization}

\begin{figure}[t]
\centering
    \includegraphics[width=\linewidth,trim= 0 5.2cm .2cm 0, clip]{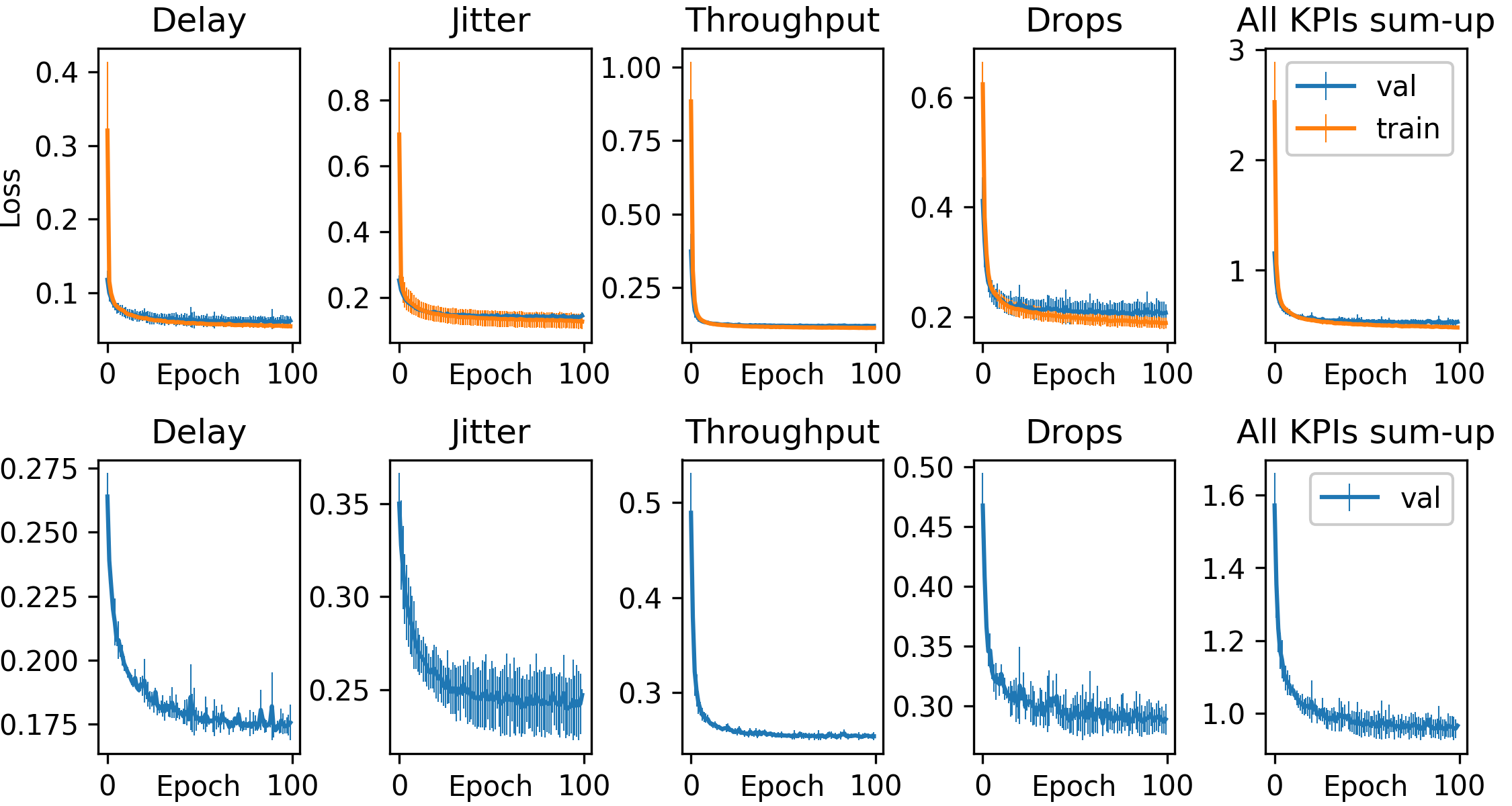}
    \vspace*{-1.5em}
    \caption{
    GLANCE's MTL training and validation losses converge well over time with continuous traffic input in NSFNET. 
    }
    \label{f:cont_lc}
\end{figure}

We set up this experiment in the fixed-flows NSFNet scenario.
The goal of load optimization is to find the traffic, i.e., $\bbtau{\,\in\,}\mbR^{2F}$ vectors, whose resulting KPIs through ns-3 will be close to the target KPIs.
In addition to efficiency, another advantage of GLANCE (or other LDTs) over SB in this application lies in its differentiable nature as a neural network architecture.  
To bring this into effect, a continuous traffic space is necessary, but it was discrete with $\tau{\,\in\,}\{1,10,20\}$ in previous experiments.
Hence, in the first step, we validate GLANCE with continuous traffic inputs $\tau{\,\sim\,}\ccalU(1,20)$.  
Fig.~\ref{f:cont_lc} shows the new learning curves of training and validation losses throughout 100 completed epochs. 
In this case, GLANCE can converge smoothly and generalize well without serious under- or over-fitting issues.
We employ as our chosen evaluator the model in the first CV fold at the epoch with minimum validation loss.
This model has an average NMAE (over all KPIs) of $0.237$ on the test set, significantly lower than SimBase$^{+\!+}$'s $0.260$.
The validation of GLANCE's generalizability to continuous traffic sets the stage for using it as a differentiable network evaluator.

\begin{figure}[t]
\centering
\begin{subfigure}[b]{\linewidth}
\centering
    \includegraphics[
        width=.7\linewidth,
        trim=.3cm .3cm .3cm .4cm
    ]{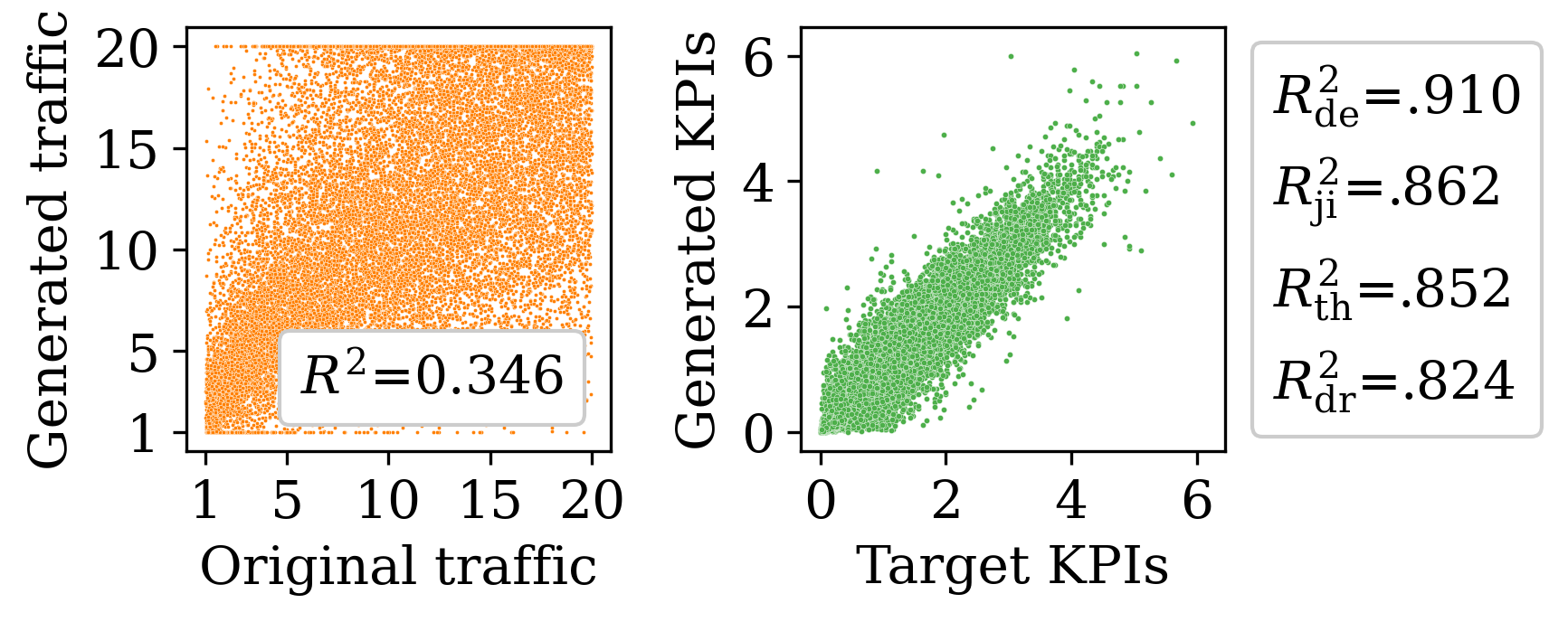}
    \caption{
    Left: generated versus original traffic.
    Right: generated versus target KPI values.
    The annotated $R^2$ values show that our generation fits the original traffic poorly but the target KPIs well.
    }\label{f:gen_traf_a}
    \vspace*{.5em}
\end{subfigure}
\vspace*{.5em}
\begin{subfigure}[b]{\linewidth}
\centering
    \includegraphics[width=.85\linewidth,trim= 0 .2cm 0 0, clip]{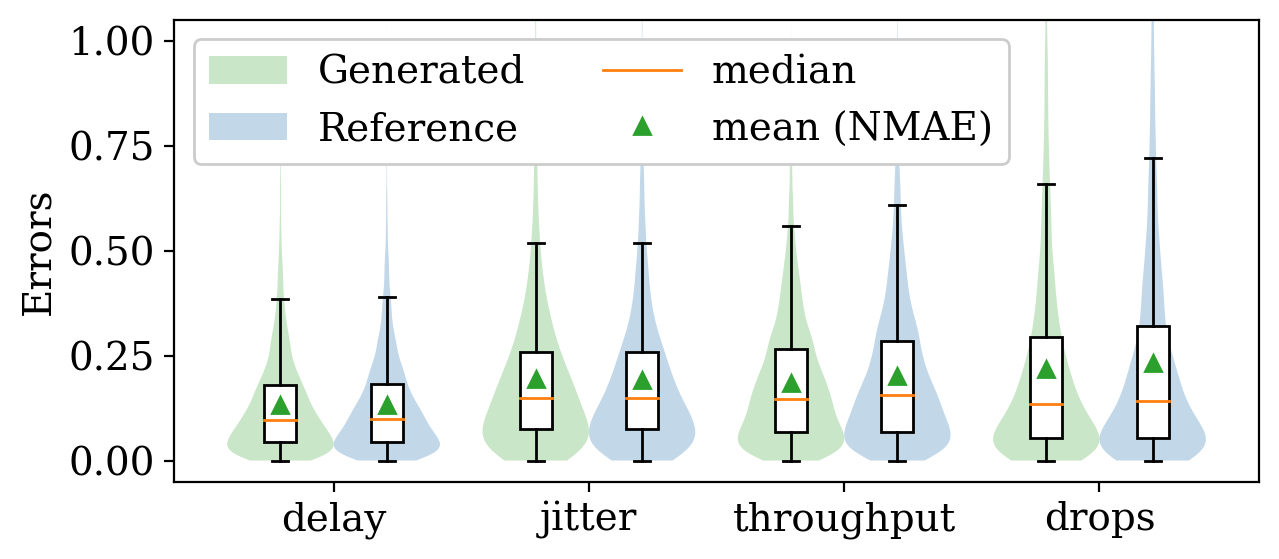}
    \caption{
    Comparison between generated and benchmark KPIs on their absolute errors relative to target KPIs.
    The shades show the error distributions, with median and mean error values annotated on the box plots.
    }\label{f:gen_traf_b}
\end{subfigure}
\vspace{-6mm}
\caption{
    Visualizing results for traffic load management in NSFNET.
}\label{f:gen_traf}
\end{figure}

Given target 
$\bbk_\targ$, let $\hat{\bbtau}{\,=\,}G_\traf(\bbk_\targ|\Psi_\tau)$ denote the generation of traffic following the gradient-based approach detailed in Section~\ref{s:alg}, where $\Psi_\tau$ represents the trained GLANCE model parameterized by $\bbtau$. 
While direct evaluation could be done in the input space by comparing the generated $\hat{\bbtau}$ with the original $\bbtau$, it is not advisable to do so due to the {non-invertible} nature of this problem. 
The left plot of Fig.~\ref{f:gen_traf_a} illustrates this point by displaying $\hat{\bbtau}$ against $\bbtau$ for 1000 instances, revealing a poor fit with a low coefficient of determination ($R^2$) value.
Instead, we rely on ns-3 once more to simulate and compare the simulated $\hbk_{\gen}$ with the original target $\bbk_{\targ}$ in the KPI space.
In the right plot of Fig.~\ref{f:gen_traf_a}, an overall $R^2$ of $0.886$ is achieved between $\hbk_{\gen}$ and $\bbk_{\targ}$, which affirms the effective role GLANCE plays herein.

Fig.~\ref{f:gen_traf_b} extends the comparison between $\varepsilon_\gen$ and $\varepsilon_\mathrm{bm}$ by showing the distribution, median, and mean values of the absolute errors for individual KPI types. 
Clearly, the performance of generated KPIs is comparable to that of the benchmark KPIs, which indicates successful traffic management. 
Among the four KPIs, the best performing one being delay aligns GLANCE's learning performance, which excelled in delay as well (see Fig.~\ref{f:cont_lc}).
This observation underscores the importance of using an accurate network evaluator for downstream tasks.
Utilizing the transfer learning strategy on GLANCE, we can uphold the evaluation performance with minimal retraining efforts in situations where input data distribution constantly deviates.

The application of traffic load management holds significant practical value.
In a smart city, sensors at intersections collect and exchange data on vehicle movements and road conditions to enhance transportation efficiency. 
The transmission paths for this data are predetermined based on the communication infrastructure and range limitations.
However, given the fluctuating volume of data generated at each intersection throughout the day, rescheduling the transmission load becomes essential to maintain unaffected performance and avoid wasting resources.
Our proposed framework can ensure that the system continues to function efficiently despite varying data loads.

\subsubsection{Flow destination optimization}

\begin{figure}[t]
\begin{subfigure}[b]{\linewidth}
\centering
    \includegraphics[
        width=\linewidth,
        trim= .35cm .4cm .3cm .5cm
    ]{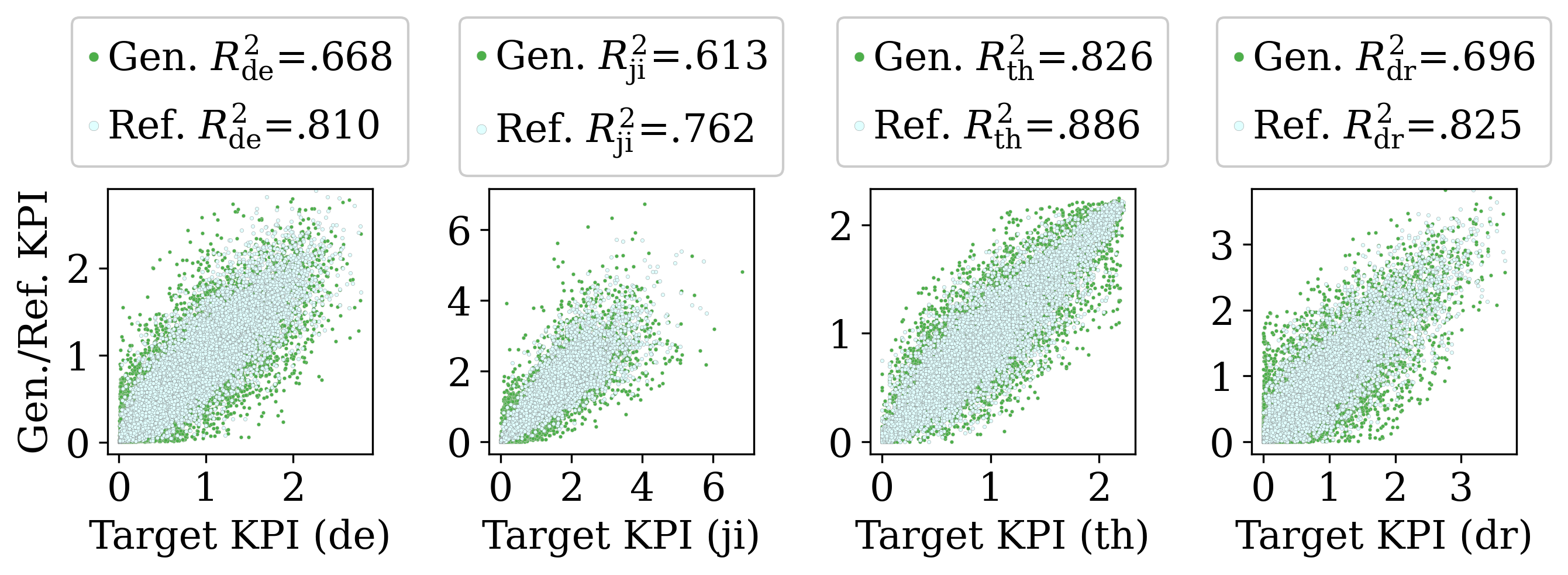}
    \vspace*{-1em}
    \caption{
    Comparison between generated and benchmark KPIs' goodness of fit against target KPIs for each KPI type.
    }\label{f:offload_kpi_err_a}
    \vspace*{.5em}
\end{subfigure}
\begin{subfigure}[b]{\linewidth}
\centering
    \includegraphics[width=.85\linewidth,trim= 0 .2cm 0 0, clip]{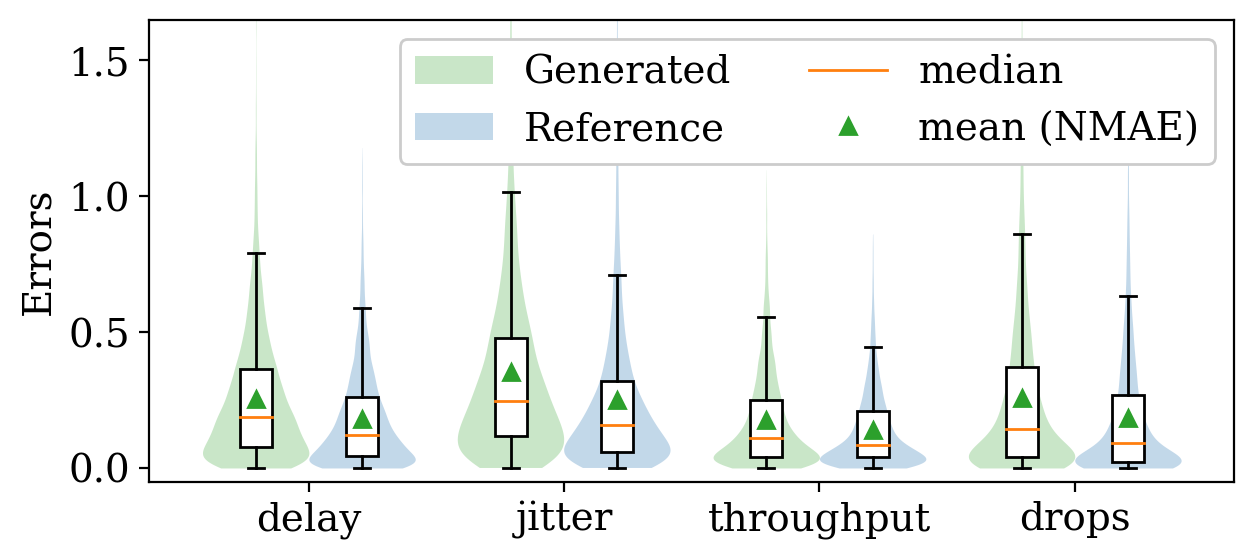}
    \caption{
   Comparison of generated and benchmark KPIs' absolute errors relative to target KPIs for each KPI type.
    }\label{f:offload_kpi_err_b}
\end{subfigure}
\caption{
Results and performance visualized for flow destination optimization in RegGrid.
}\label{f:offload_kpi_err}
\end{figure}

This experiment is configured in the RegGrid scenario, where $\bbf_\src{\,\in\,}\ccalN^{F}$ (source node indices) of flows are fixed and  $\bbf_\dst{\,\in\,}\ccalN^{F}$ (destination node indices) are to be determined.
The goal remains the same: to manage the KPIs of the flows generated.
The major difference herein is our selection of a hill-climbing algorithm instead of a gradient-based one for optimization steps. 
This follows from the fact that the destination nodes have to be chosen from a discrete set.
Previous experiments in Section~\ref{s:exp:gen:flow} suggest that GLANCE faces greater difficulties in predicting network KPIs when there are changes in the input flows.
This is evident from the increased errors observed -- compared to that of the fixed-flows in \mbox{Fig.~\ref{f:tabfig}b} -- in the random-flows \mbox{Fig.~\ref{f:tabfig}c}, where a significant variation in performance is also present across different KPI types, with throughput leading and jitter trailing. 
Based on these factors, we refine our visualization for the flow optimization results in Fig.~\ref{f:offload_kpi_err}.
Specifically in Fig.~\ref{f:offload_kpi_err_a}, given the difficulty in visualizing the current input space, we shift focus to the KPI space and plot the generated versus target KPIs separately for each type to avoid overlooking insights tied to individual types.
Similarly to the previous case, Fig.~\ref{f:offload_kpi_err_b} visualizes the generated and benchmark errors' statistics.  
In this case, the performance gaps between generated and benchmark KPIs are not surprising, given our understanding that GLANCE's performance can be affected when it needs to generalize to more entities.
Whether or not this performance is satisfactory depends on the specific problem and objective in practice.

Alternative to the MAE objective used thus far, hinge loss may also come into interest if one desires to ensure a minimal quality of service (QoS) instead of symmetrically approximating a specific KPI profile.
In this scenario, we define failed cases as where the generated KPI violates the target KPI bound, i.e., $\hat{k}_\gen{\,>\,}k_\targ$ for delay, jitter, drops, or $\hat{k}_\gen{\,<\,}k_\targ$ for throughput.
By this metric, the failure ratios amount to $7.2\%$ for delay, $5.3\%$ for jitter, $16.8\%$ for throughput, and $9.3\%$ for drops. 
It is interesting to notice that the ordering of performance on each KPI type by this objective is converse to that by absolute errors. 
This discrepancy can likely be attributed to GLANCE being trained with a symmetric loss function.
The mismatch between the network evaluation and management objectives may interfere with the final performance. 
Therefore, it is recommended to maintain consistency (by retraining or finetuning the LDT) in subsequent tasks to achieve optimal and anticipated performance.
In summary, what is demonstrated here may serve as a starting point for a wide range of applications centered around computational offloading, such as mobile edge computing, wireless federated learning, and autonomous vehicular networks, among others.

\section{Conclusions}\label{s:conclusion}

As a learnable digital twin to the network simulator \mbox{ns-3}, GLANCE exhibited powerful predictive capabilities via different supervised training paradigms.
Its relational node-net implicitly captures interference between nodes, which is critical information for wireless network performance. 
Its link-net and recurrent path-net can integrate link embeddings in the correct order.
Together, GLANCE can learn universal path embeddings representing inherent structures of network topology and flows, reflected in its strong generalizability against these entities.
Example traffic management applications held promise for enhanced efficiency and effectiveness by leveraging GLANCE in diverse network environments. 
In future research, we will further explore the integration of GLANCE into the RL framework for computational offloading. 
By combining RL's adaptive learning capabilities with GLANCE's comprehensive understanding of network structures, we anticipate significant advancements in computational offloading techniques.

\appendices
\section{}\label{ap:notations}
The following Table~\ref{tab:notation} summarizes important notation used in this paper. 
Generally speaking, spaces and sets are typeset in calligraphic letters (e.g., $\ccalA$).
Vectors are in bold lowercase (e.g., $\bba$) and matrices in bold uppercase (e.g., $\bbA$).
Alternative notations for vectors $[a_i]^{n-1}_{i=0}{\,=\,}[a_0,...,a_{n-1}]$ and sets $\{a_i\}^{n-1}_{i=0}{\,=\,}\{a_0,...,a_{n-1}\}$ may be present for clarification.
The notation $f(x;\bbW)$ represents a function $f$ parameterized by $\bbW$.
In addition to defining vectors or lists, we also use brackets for indexing, e.g., $\bbx[i]$ is the $i^\text{th}$ element of $\bbx$, sometimes simplified as $x_i$.
Apart from defining sets, we also use braces in subscripts to simplify notations, e.g., `$x_{\{a,b,c\}}$' is short for `$x_a$, $x_b$, and $x_c$'.
Topologically, $L_\mathrm{out}(n)$ denotes the set of links emanating from node $n$. 
In our network system, values of configurations and outcomes, such as traffic and KPIs, are non-negative by default.

\begin{spacing}{.95}
{\small
\begin{longtblr}[
  caption = Important notation.,
  label = {tab:notation},
]{
  colspec = {rp{180pt}},
  rowhead = 1,
} 
\toprule
Notation & Description\\
\hline
$\ccalN, \ccalL, \bbA$    &  Node space, link space, and adjacency matrix. Network topology is defined as $\ccalG{\,=\,}(\ccalN, \ccalL, \bbA)$.\\
$\ccalF$    & ${=\,}\{(s_i,d_i)\}_{i=0}^{F-1}$ denotes the multi-flow space of $F$ unique flows, with $\bbf_i{\,=\,}(s_i,d_i)$ being the $i^\text{th}$ flow.\\
$\bbf_\src$  & ${=\,}[s_i]_{i=0}^{F-1}$, the index vector of all source nodes. For the destination-node counterpart, $\bbf_\dst{\,=\,}[d_i]_{i=0}^{F-1}$.\\
$P(\ccalF)$, or $\bsP$ & Correspondent paths, with {$\bbp$ denoting a single path and $p_i$ the $i^\text{th}$ link in that path}.\\
$t_\onoff$   & $\sim\ccalT(\tau_\onoff)$, sampling on- and off-traffic random variables from \mbox{$\tau_\on$- and $\tau_\off$-parameter} random distributions, respectively, in \mbox{ns-3}. \\
$\bbtau_\onoff$  &  ${=\,}[\tau_{\onoff}^i]_{i=0}^{F-1}$ represent the traffic vectors concerning the on and off time distributions, respectively. A zig-zag flattened version can be expressed as $\bbtau{\,=\,}\left[\tau_{\on}^{0}, \tau_{\off}^{0},\cdots, \tau_{\on}^{F-1}, \tau_{\off}^{F-1}\right]{\,\in\,}\mbR^{2F}$.\\
$\ccalX$    &Network configuration space. 
More specifically, $\ccalX{\,=\,}\{\ccalX_g, \ccalX_t, \ccalX_f\}$, in which the subspaces represent topology, traffic, and flows, respectively, and constitute the input space for this study, where\\
$\ccalX_f$ & $=\,$\(\left\{\begin{array}{rl}
\ccalF  & \text{for \mbox{ns-3},} \\
P(\ccalF)  & \text{for LDTs.} \end{array} \right.\)\\
$\bbx$    &${\in\,}\ccalX$ denotes an input instance.\\
$\ccalK$    &${=\,}\{\mathrm{de(lay), ji(tter), th(roughput), dr(ops)}\}$, output space of the network system, or the $K{\,=\,}4$ KPIs on which this work is focused. \\
\bottomrule
\end{longtblr}
}
\end{spacing}

\bibliographystyle{IEEEbib}
{
    \linespread{.95}
    \footnotesize
    \bibliography{IEEEabrv,references}
}

\end{document}